\documentclass[twocolumn,aps,prl,superscriptaddress,english]{revtex4-2}
\usepackage{amsmath,amsfonts, amssymb, amsthm, dsfont}
\usepackage{yfonts}
\usepackage{bm}
\usepackage{mathrsfs}
\usepackage{graphicx}
\usepackage{verbatim}
\usepackage[colorlinks=true,citecolor=blue]{hyperref}
\usepackage[normalem]{ulem}
\usepackage{empheq}
\usepackage{array}
\newcolumntype{C}[2]{>{\centering\arraybackslash}m{#1}}
\usepackage{tensor}
\usepackage{tikz}
\usepackage{comment}
\usepackage{float}
\usetikzlibrary{calc}
\usepackage{mathtools}
\usepackage{adjustbox}
\usepackage{newfloat}
\DeclareFloatingEnvironment[name={Figure}]{suppfigure}
\makeatletter
\def\maketitle{
\@author@finish
\title@column\titleblock@produce
\suppressfloats[t]}
\makeatother

\newcommand{\ee}{\mathrm{e}} 
\newcommand{\ii}{\operatorname{i}} 
\newcommand{\VV}{\mathcal{V}}
\newcommand{\ZZ}{\mathbb{Z}} 
\newcommand{\HH}{\mathbb{H}} 
\newcommand{\sgn}{\mathrm{sgn}}
\newcommand{\Tr}{\mathrm{Tr}}

\newcommand{\dg}{\dagger}

\newcommand{\ket}[1]{\lvert#1\rangle}
\newcommand{\bra}[1]{\langle#1\rvert}

\newcommand{\lrangle}[1]{\langle#1\rangle}
\newcommand{\abs}[1]{\left\lvert#1\right\rvert}

\tikzset{
my dash/.style={dash pattern=on 1 off 0.5
                }
         }

\newcommand{\rag}{\rangle}
\newcommand{\lag}{\langle}
\hypersetup{colorlinks=true}

\begin{document}

\title{Measurement protocol for detecting correlated topological insulators in synthetic quantum systems}
\author{Yixin Ma}
\affiliation{Kavli Institute for Theoretical Sciences, University of Chinese Academy of Sciences, Beijing 100190, China}
\author{Chao Xu}
\email{xu-chao@tsinghua.edu.cn}
\affiliation{Institute for Advanced Study, Tsinghua University, Beijing 100084, China}

\author{Shenghan Jiang}
\email{jiangsh@ucas.ac.cn}
\affiliation{Kavli Institute for Theoretical Sciences, University of Chinese Academy of Sciences, Beijing 100190, China}

\date{\today}
\begin{abstract}
Two-dimensional topological insulators, characterized by symmetry-protected anomalous boundary modes, have been generalized to the strongly correlated regime for both bosonic and fermionic systems.
As correlated topological insulators~(TI) approach experimental realization in quantum simulators, conventional probes, such as transport measurements, are not easily applicable to these synthetic platforms.
In this study, we focus on two examples of correlated TI: a bosonic TI protected by $\ZZ_2\times U(1)$ symmetry and the fermionic quantum spin Hall insulator protected by time-reversal symmetry. 
We propose a unified, readily implementable protocol based on measuring the disorder parameter $\lrangle{\exp(\ii\theta \hat{Q}_A)}$ for a large subregion $A$, with $\hat{Q}_A$ the total charge operator within $A$.
Our key finding is that this quantity exhibits non-analytical dependence on $\theta$ for correlated TI, a signature robust against decoherence.
We establish this diagnostic through both numerical simulations and analytical derivations.
This protocol is well-suited for implementation on near-term quantum simulation platforms, providing a direct route to experimentally confirm correlated TI.
\end{abstract}

\maketitle

\emph{Introduction.--}
The preparation and detection of topological quantum phases~\cite{pichler2016measure,andersen2019entanglement,Grusdt2016,wen2017} have been substantial topics in condensed matter physics over the past decades.
Famous examples include two-dimensional topological insulators~\cite{KaneMele2005,bernevig2006quantum,HasanKane2010,QiZhang2011,chang2013experimental}, characterized by trivial bulk but helical metallic edge modes protected by charge conservation and time-reversal symmetry. 
Although first proposed as electronic band insulators, such phases have been extended to strongly correlated boson and fermion systems~\cite{wangBosonicAnomaliesInduced2015,wang2023exactly}, which we referred as ``correlated topological insulators''~(correlated TI), a class of symmetry-protected topological~(SPT) phases~\cite{GuWen2009,ChenGuWen2011,chenSymmetryProtectedTopological2013,senthil2015symmetry,wang2023exactly} where protecting symmetry includes $U(1)$.

Recently, the experimental realization of interacting topological phases in synthetic quantum platforms, including optical lattices~\cite{goldman2010,goldman2016,schafer2020tools}, trapped ions~\cite{debnath2016demonstration,friis2018observation,zhang2017observation,wright2019benchmarking, iqbal2024non,iqbal2024topological}, superconducting quantum processors~\cite{satzinger2021realizing,clarke2008superconducting,reagor2018demonstration}, and Rydberg atom tweezers arrays~\cite{deleseleuc2019,ebadi2021,semeghini2021probing}, has become a subject of intense interest. 
Detection of topological phases in these platforms presents unique challenges compared to those encountered in weakly interacting topological materials. 
For topological insulators, traditional experimental methods, such as transport measurements~\cite{bernevig2006quantum,fei2017edge} and scanning tunneling microscopy~\cite{tang2017quantum} are not directly applicable to synthetic quantum systems.
Furthermore, interactions may gap out the metallic edge modes~\cite{WuBernevigZhang2006,XuMoore2006,Metlitski20191d,sullivan2020interacting,ning2021edge}, necessitating novel measurement schemes based on bulk topological properties. 

Unlike symmetry-breaking phases characterized by local order parameters, the bulk measurement of topological phases usually require non-local probes -- such as topological entanglement entropy in 2D topological orders~\cite{kitaev2006,levin2006detecting} or string order parameters in 1D SPT phases~\cite{perez2008string,pollmann2012}.
While challenging to measure in materials, synthetic quantum platforms enable direct access to many such quantities, for instance, string order parameters for the Haldane phase~\cite{sompet2022a}, many-body topological invariants in 1D Su-Schrieffer-Heeger chains~\cite{elben2020}, and many-body Chern numbers in 2D Chern insulators~\cite{cian2021}. 
However, generalizing the measurement protocol to include 2D correlated TI remains an open challenge.

\begin{figure}
    \centering
    \includegraphics[scale=0.65]{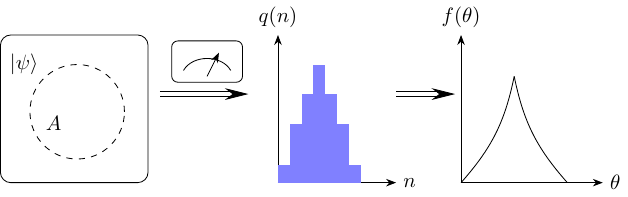}
    \caption{
        Detection scheme for disorder parameters in quantum simulators.
        The protocol involves performing $N_M$ projective measurements of the many-body wavefunction $\ket{\psi}$ in the particle number basis for a sub-region $A$.
        This yields the distribution $q(n)$, counting the occurrences of particle number $n$ within $A$.
        The scaled disorder parameter $f(\theta)\equiv-{\abs{\partial A}^{-1}}\cdot \ln\abs{U_A(\theta)}$ is then computed from $q(n)$ via post data-processing. 
        A characteristic non-smooth behavior of $f(\theta)$ serves as a direct signature of a nontrivial correlated TI.
    }
    \label{fig:scheme_sketch}
\end{figure}

\begin{figure*}
    \centering
    \includegraphics[width=0.9\linewidth]{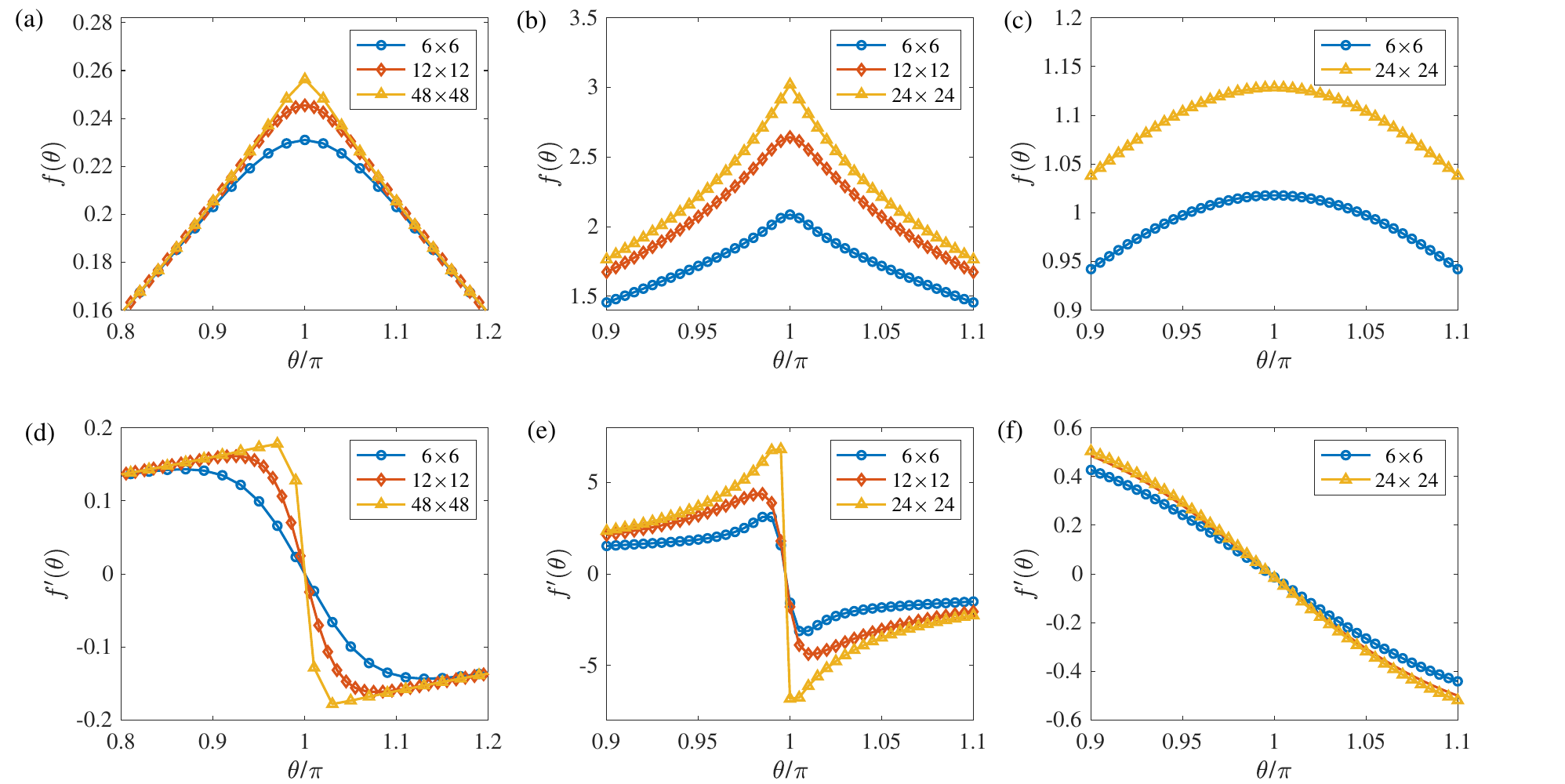}
    \caption{
    Numerical results for the disorder parameter.
    (a,d) Bosonic correlated TI.
    (b,e) Fermionic QSH insulator.
    (c,f) Trivial fermionic insulator with time reversal symmetry.
    Top row (a-c): scaled disorder parameter $f(\theta)\equiv -\abs{\partial A}^{-1} \ln\langle{U_A(\theta)}\rangle$.
    Bottom row (d-f): its derivative $f'(\theta)$.
    Different subsystem sizes $\abs{\partial A}$ are indicated in the legends.
    The discontinuity behavior of $f'(\theta)$ at $\theta=\pi$ in (d,e) provides a direct signature of correlated TI.
    }
    \label{fig:num_resutls}
\end{figure*}

In this work, we propose a novel measurement scheme for identifying two-dimensional correlated TI by evaluating the \emph{disorder parameters}\cite{wang2024distinguishing,chenTopologicalDisorderParameter2022,cai2024,liu2014,jiang2023,cai2024disorder,mao2025probing} $\langle U_A(\theta)\rangle\equiv \bra{\psi}\exp(\ii\theta \hat{Q}_A)\ket{\psi}$ on a single ground state wavefunction $\ket{\psi}$.
In particular, we focus on two examples: the $\ZZ_2$ Ising symmetric bosonic insulator and the time-reversal invariant fermionic insulator.
For these insulating phases, the disorder parameter exhibit an area law scaling: 
\begin{equation}
-\ln\abs{\lrangle{U(\theta)}}=f(\theta) \abs{\partial A}+\cdots
\label{eq:area_law_disorder_param}
\end{equation}
where $\abs{\partial A}$ is the boundary length of $A$.
Our central finding is that correlated TI are distinguished by the universal non-analytic features in $f(\theta)$, while trivial insulators generally exhibit a smooth $\theta$-dependence.

This protocol is well-suited for implementation on synthetic quantum systems, requiring only \emph{single-state measurement} of particle numbers within region $A$ -- a capability already demonstrated in various quantum simulation platforms~\cite{sherson2010single,yao2023observation}, and $f(\theta)$ can be obtained through post-processing of the measured particle number data, as demonstrated in Fig.~\ref{fig:scheme_sketch}.
Further, as we will analyze later, such a singular signature of $f(\theta)$ exhibits robustness against environment noises, enhancing its experimental viability.

This paper is structured as follows.
We begin by presenting numerical evidence for the proposed diagnostic for various models of correlated insulators.
Subsequently, we provide an analytical derivation of the function $f(\theta)$ based on the reduced density matrix~(RDM) formalism.
Finally, we discuss potential challenges and considerations relevant to experiment.

\begin{figure}
    \centering
    \includegraphics[scale=0.6]{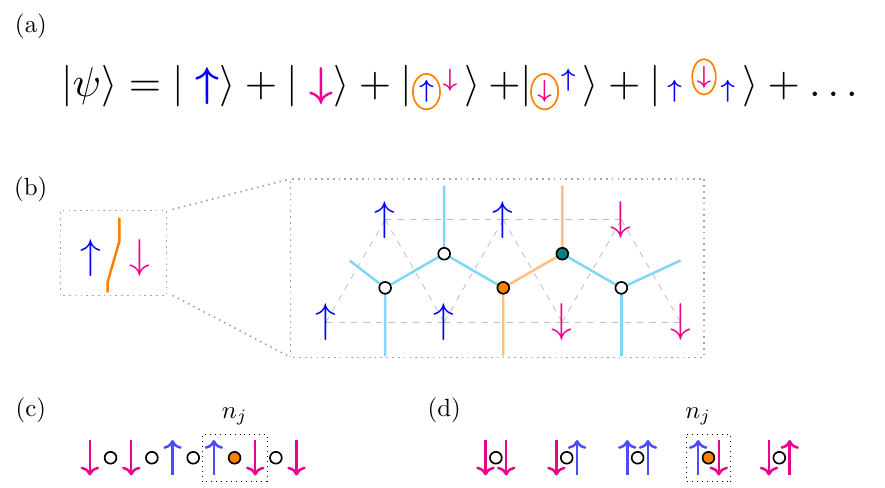}
    \caption{
    Bulk fixed-point wavefunction and boundary theory for the $U(1)\times \mathbb{Z}_2$ bosonic correlated TI.
    (a) The fixed-point wavefunction is an equal-weight superposition of all allowed configurations.
    (b) The decoration pattern of a typical configuration.
    Here, the orange/teal point refers to a positive/negative charged boson.
    (c) A typical boundary configuration. Here, the boundary $U(1)$ symmetry action is non-onsite.
    (d) An extended boundary Hilbert space, where each site holds two Ising spins. 
    The symmetry action is onsite in this extended Hilbert space. 
    }
    \label{fig:boson_wf}
\end{figure}

\emph{Numerical results.--}
Numerical evidence for the non-smooth behavior of $f(\theta)$ is presented in in Fig.~\ref{fig:num_resutls} for both bosonic and fermionic TI.
For the bosonic correlated TI with $U(1)\times\ZZ_2$ symmetry, we study a fixed point wavefunction  of this phase~\cite{horinouchi2020solvablelatticemodel21d,wang2022exactlyu1set}.
As illustrated in Fig.~\ref{fig:boson_wf} (a,b), the model consists of $\ZZ_2$ Ising spins on a triangular lattice, and $U(1)$-charged hard-core bosons on the dual honeycomb lattice.
At the fixed point, the wave function is an equal weight superposition of all $\ZZ_2$ domain wall configurations on the dual honeycomb lattice, whose vertices are decorated with positive/negative charged bosons on the $u/v$ sub-lattices, respectively.
A detailed analysis of the symmetry and edge anomaly of this wavefunction, using PEPS formalism, is provided in supplemental section II~\cite{SM}.
 
We compute $f(\theta)$ and its derivative use the Tensor Network Renormalization method~\cite{evenbly2015tensor}, with maximum cutoff $D_{c}=64$.
The results, shown in Fig.~\ref{fig:num_resutls}(a,d), clearly demonstrate that $f'(\theta)$ develops an increasingly sharp discontinuity at $\theta=\pi$ as the size of region $A$ increases. 

The fermionic system we study is the celebrated Kane-Mele model~\cite{KaneMele2005} on the honeycomb lattice: 
\begin{align}
    H_{KM}&=\sum_{\lag ij \rag}c^{\dag}_ic_j+it\sum_{\lag\lag ij\rag\rag}\nu_{ij}c^{\dag}_i\sigma_z c_j+m\sum_{i}c^{\dag}_i\tau_z c_i
\end{align}
where $\nu_{ij}=\sgn(\mathbf{d}_1\times \mathbf{d}_2)_{z}$ with $\mathbf{d}_1$ and $\mathbf{d}_{2}$ unit vectors along the two bonds of $\mathrm{r}_{ij}$.  
We compute $f(\theta)$ for both quantum spin Hall~(QSH) phase ($t=m=0.02$) and trivial insulator ($m=20t=0.4$) at half-filling.
The results, shown in Fig.~\ref{fig:num_resutls}(b,e), reveal that in the QSH insulator, $f(\theta)$ exhibits non-smooth behavior with a singular derivative at $\theta=\pi$.
In contrast, the trivial phase, depicted in Fig.~\ref{fig:num_resutls}(c,f), shows smooth $\theta$-dependence.
Although our calculations are performed on the non-interacting case, we emphasis that the singular behavior of $f(\theta)$ remains robust against interactions, which is supported by our analytical derivation for the correlated QSH phase, with details provided in the supplemental section II and IV.

\emph{Disorder parameters from RDM.--} 
The analytical derivation for the disorder parameter is most naturally formulated using RDM.
Our key observation is that as for a gapped ground state, the entanglement of region $A$ with the rest of the system is concentrated on the 1D boundary $\partial A$, this allows representing the RDM by \emph{a one-dimensional mixed state} $\rho_{\partial A}$ defined on $\partial A$.
$\lrangle{U_A(\theta)}$ then reduces to evaluating a specific expectation value in this 1D system.
Crucially, symmetry action on $\rho_{\partial A}$ is anomalous, which results in the non-smooth behavior in $f(\theta)$. 

We begin with the RDM of region $A$ 
\begin{equation}
    \begin{aligned}
        \rho_{A}=\Tr_{\bar{A}}\ket{\psi}\bra{\psi}
        =\sum_{\alpha=1}^{\chi_A}e^{-\lambda_\alpha}|\phi_A^\alpha\rag\lag\phi_A^\alpha|.
    \end{aligned}
    \label{eq:psi_decomp}
\end{equation}
where $\{\ket{\phi^\alpha}\}$ are orthonormal Schmidt states defined in Hilbert space $\HH_A$, and $\{\lambda_\alpha\}$ constitute the entanglement spectra.
The Schmidt indices $\{\alpha\}$ define a $\chi_A$-dimensional \textit{entanglement Hilbert space}, denoted as $\HH_{\partial A}$, isomorphic to $\mathrm{span}\{\ket{\phi^\alpha_A}|\alpha=1,\dots\chi_A\}$.
For the case where $\ket{\psi}$ is represented by a projected entangled pair state~(PEPS), $\HH_{\partial A}$ is identified as virtual legs at the boundary. 
More generally, for a non-chiral gapped phase, we assume that $\HH_{\partial A}$ can be well approximated by the Hilbert space of a 1D chain at $\partial A$: $\HH_{\partial A}=\otimes_{j\in \partial A}\HH_j$.
\begin{figure}
    \includegraphics[scale=0.65]{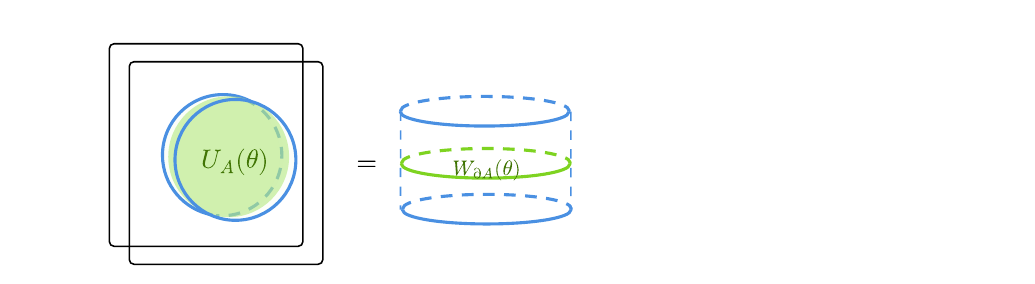}
    \caption{Calculating disorder parameter with RDM defined at $\partial A$. After the isometry $U$, the RDM $\rho_A$ is mapped to $\rho_{\partial A}$ defined on the entanglement space. The disorder parameter $\lag U_A(\theta)\rag$ can be computed as the expectation value of $W_{\partial A}(\theta)$ in the 1D mixed state $\rho_{\partial A}$.}
    \label{fig:dis}
\end{figure}

Let $U$ be the isometry from $\HH_{\partial A}$ to $\HH_A$, we then define a 1D mixed state $\rho_{\partial A}$ supported on $\HH_{\partial A}$ as
\begin{equation}
    \rho_{\partial A}=U^\dg\cdot \rho_{A}\cdot U
\end{equation}
The global symmetry $U(g)$ induces symmetry $U_A(g)$ on $\rho_A$, with $[\rho_{A},U_{A}(g)]=0$, which further imposes symmetry $W_{\partial A}(g)\equiv U^\dagger\cdot U_{A}(g)\cdot U$ on $\rho_{\partial A}$~(see details in supplemental section I):
\begin{equation}
    \rho_{\partial A}=W_{\partial A}(g)\cdot \rho_{\partial A}\cdot W_{\partial A}^\dg(g)
    \label{eq:bdry_rdm_weak_sym}
\end{equation}
As illustrated in Fig.~\ref{fig:dis}, it enables the calculation of the disorder parameter entirely within $\HH_{\partial A}$:
\begin{align}
   \lag U_{A}(\theta)\rag= \Tr[\rho_{\partial A} W_{\partial A}(\theta)]
    \label{eq:dis_order_boundary_leg}
\end{align} 

\emph{The anomalous symmetry on RDM.--}
Physically, $W_{\partial A}(g)$ are identified as the anomalous boundary symmetry, encoding the bulk topological data.
While $W_{\partial A}$ admits a natural matrix product unitary operator (MPUO) representation, reflecting its non-onsite nature~\cite{chen2011two,williamson2016matrix}, we demonstrate~(see supplemental section II) that it can be equivalently expressed as an onsite action $W_{\partial A}(g)=\prod_j W_j(g)$.
This is achieved in two key steps~\cite{wen2023bulk,jiangAnyonCondensationGeneric2017,xu2024unveiling}:
\begin{enumerate}
    \item  Enlarging the entanglement space to accommodate an extended symmetry group with on-site representation;
    \item Projecting back to the original entanglement space, and recovering the non-onsite MPUO form.
\end{enumerate}
As the on-site formulation is more tractable for the analytical derivation, we will use it henceforth.

To illustrate the above procedure, we examine the $\ZZ_2$ symmetric bosonic correlated TI as an example.
For the simplest case, its boundary Hilbert space can be modeled as a 1D Ising chain with spins are at bond centers $j+\frac{1}{2}$, with symmetries defined as follows:
\begin{itemize}
    \item The $\ZZ_2$ generator $h$ is represented as $W_{\partial A}(h)=\prod_j X_{j+\frac{1}{2}}$.
    \item The $U(1)$ action is non-onsite.
    As shown in Fig.~\ref{fig:boson_wf}(c), a unit $U(1)$ charge is assigned to the $\uparrow\downarrow$ domain wall, while all other configurations ($\uparrow\uparrow$, $\downarrow\downarrow$, and $\downarrow\uparrow$) are charge-neutral. 
    This is implemented by action $W_{\partial A}(\theta)=\exp(\ii\theta \sum n_{j})$ with $n_{j}=\frac{1}{4}(1+Z_{j-\frac{1}{2}})(1-Z_{j+\frac{1}{2}})$. 
    Here, $\sum_j n_j=N_{\uparrow\downarrow}$ counts the number of $\uparrow\downarrow$ configurations.
\end{itemize}
To resolve the non-onsite character, we embed the system into a 1D chain with \emph{two Ising spins} per site: $\ket{s^{l}_j}$ and $\ket{s^{r}_j}$ ($s=\uparrow/\downarrow$), as illustrated in Fig.~\ref{fig:boson_wf}(d).
The original bond spins are recovered by enforcing $m_j^{r}=m_{j+1}^{l}$ with $m=\ket{\downarrow}\bra{\downarrow}$.
In this enlarged Hilbert space, the symmetry action is fully onsite: $n_j=\frac{1}{4}(1+Z_{j}^{l})(1-Z_{j}^{r})$, and $W_j(h)=X_j^{l}X_j^{r}$.
Projecting back to the subspace where $m_j^{r}=m_{j+1}^{l}$ recovers the anomalous non-onsite symmetry mentioned above.
Crucially, the onsite actions do not form $U(1)\times\ZZ_2$ group, instead, they satisfy the following algebraic relations:
\begin{equation}
\begin{aligned}
    &W_j(h)\cdot n_j\cdot W_j^{-1}(h)=n_j+m_j^{l}-m_j^{r}\\
    &W_j(h)\cdot m_j^{l/r}\cdot W_j^{-1}(h)=\hat{1}-m_j^{l/r}
\end{aligned}
    \label{eq:bTI_tensor_eq}
\end{equation}
We here mention that our approach does not restrict to the Ising chain case: any 1D system satisfying Eq.~\eqref{eq:bTI_tensor_eq} together with the constraint $m_j^r=m_{j+1}^l$ exhibits the same boundary anomaly characterized by a 3-cohomology, which we will show in supplemental section III, 

Equipped with the anomalous symmetries, we are now ready to list conditions for $\rho_{\partial A}$ of the $\ZZ_2$ Ising symmetric correlated TI:
\begin{enumerate}
    \item {\bf Weak symmetries:} $\rho_{\partial A}$ obeys Eq.~\eqref{eq:bdry_rdm_weak_sym}, with onsite symmetries satisfying Eq.~\eqref{eq:bTI_tensor_eq}.
    \item {\bf Local constraints:}  The constraint $m_j^{r}=m_{j+1}^{l}$ implies that 
        \begin{equation}
        (m_j^{r}-m_{j+1}^l)\cdot \rho_{\partial A}= \rho_{\partial A}\cdot (m_j^{r}-m_{j+1}^l)=0~,~\forall j\in \partial A
        \label{eq:bdry_rdm_strong_plq_sym}
        \end{equation}
    \item  {\bf Short-range correlations:} As $\ket{\psi}$ is gapped symmetric, $\rho_{\partial A}$ should be short-range correlated: for any local operator $O$
        \begin{equation}
            \Tr(\rho_{\partial A}\cdot O^\dg_j O_{j'})-\abs{\Tr(\rho_{\partial A}\cdot O_j)}^2\sim \exp\left( \abs{j-j'}/\xi \right)
            \label{eq:bdry_rdm_src}
        \end{equation}
\end{enumerate}
These three conditions in fact indicate that the RDM $\rho_{\partial A}$ is in an intrinsic average SPT phase\cite{ma2023c,xu2025diagnosing,SM} (see details in the supplemental section III). 
As we will see, they collectively determine the universal properties of $\rho_{\partial A}$, leading to the non-smooth behavior of $f(\theta)$.

\emph{Disorder parameter from transfer matrix spectral flow.--}
By imposing uniform condition, we expect that $\rho_{\partial A}$ can be represented by a matrix product density operator~(MPDO) with local tensor $M$ for large subregion $A$:
\begin{equation}
    \rho_{\partial A}= \adjincludegraphics[valign=c]{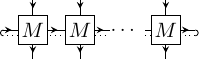}
\end{equation}
Here, the in(out)-ward arrow denotes the ket(bra) space.

As $\abs{\partial A}\gg 1$, the disorder parameter in Eq.~\eqref{eq:dis_order_boundary_leg} is dominated by the leading eigenvalue of the $\theta$-twisted transfer matrix $T(\theta)\equiv\vcenter{\hbox{\includegraphics[height=3.5em]{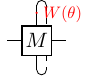}}}$, i.e., 
\begin{align}
\lrangle{U_A(\theta)}=\Tr([T(\theta)]^{\abs{\partial A}}) 
\simeq \sum_{l=1}^{k}[\lambda_{l}(\theta)]^{\abs{\partial A}}~,
\label{eq:disorder_param_transfer_mat_eigenval}
\end{align}
where $\{\lambda_l(\theta)|l=1,\ldots,k\}$ are $k$-degenerate dominant eigenvalues in terms of modulus. 
Note that due to the short-range correlated condition Eq.~\eqref{eq:bdry_rdm_src}, $k=1$ for the untwisted transfer matrix $T\equiv T(\theta=0)$.

Hence, the behavior of the disorder parameter reduces to studying \emph{the spectral flow of $T(\theta)$} as $\theta$ varies. 
Symmetry constraints on $T(\theta)$ come from symmetries of $M$.
From the weak symmetry (Eq.~\eqref{eq:bdry_rdm_weak_sym}) and local constraints (Eq.~\eqref{eq:bdry_rdm_strong_plq_sym}) condition on $\rho_{\partial A}$, we derive symmetry action $V(h)$, $V(\theta)$ and $m^{u/d}$ on the virtual space\cite{SM}, resulting in
\begin{equation}
   \adjincludegraphics[valign=c]{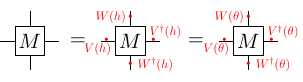}
\label{eq:tensor_weak_sym}
\end{equation}
as well as
\begin{equation}
    \adjincludegraphics[valign=c]{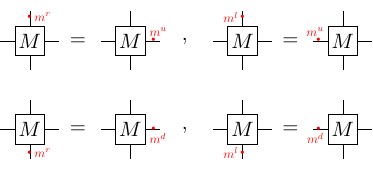}
\label{eq:tensor_strong_plq_sym}
\end{equation}
Algebraic relations of these virtual leg symmetries follow Eq.~\eqref{eq:bTI_tensor_eq}.
As detailed in supplemental section III, these relations imply a $\ZZ_2\times\ZZ_2$ symmetry generated by $\VV^{u/d}_{\theta}$ on $T(\theta)$:
\begin{equation}
 T(\theta)=\mathcal{V}^a_\theta\cdot T(\theta)\cdot \left( \mathcal{V}^a_\theta \right)^\dg~,~
  a=u,d
\end{equation}
where $\mathcal{V}^a_{\theta}\equiv V(h)\ee^{\ii\theta(m^a-\frac{1}{2})}$, satisfying $\mathcal{V}_\theta^a=-\mathcal{V}_{\theta+2\pi}^a$.

We now investigate the spectral flow of $T(\theta)$.
Let $v^r_0$ be the unique (right) dominant eigenstate of $T$, with eigenvalue $\lambda_0=1$ and quantum number $s_0\in\{1,-1\}$ under $\mathcal{V}^u_{\theta=0}$.
As $\theta$ flows 0 to $2\pi$,  continuity implies the quantum number remain fixed, $s_0(\theta)=s_0$.
However, since $\mathcal{V}_0^a=-\mathcal{V}_{2\pi}^a$, the evolved eigenvector $v_0^r(2\pi)$ must have quantum number $-s_0$ under $\mathcal{V}_0^a$,
Therefore, $v^r_0(2\pi)$ is a distinct eigenvector of $T=T(2\pi)$ from $v^r_0$ with eigenvalue $\abs{\lambda_1}<1$. 
This enforces a crossing of the magnitude of their eigenvalues, $\abs{\lambda_0(\theta)}$ and $\abs{\lambda_1(\theta)}$, as $\theta$ varies, and hence leads to non-smooth behavior of $\abs{\lrangle{U_A(\theta)}}$ according to Eq.~\eqref{eq:disorder_param_transfer_mat_eigenval}.

As detailed in supplemental section III, the Hermiticity of $\rho_{\partial A}$ further forces $\abs{\lambda_0(\pi)}=\abs{\lambda_1(\pi)}$.
The simplest scenario is the case in Fig.~\ref{fig:scheme_sketch}, where two dominant eigenvalues cross at $\theta=\pi$\footnote{Other types of singularities are possible, e.g. with more corners, see SM~\cite{SM} for more discussion.}.

The analytical framework can be extended to the QSH insulator, where a similar non-smooth behavior of $f(\theta)$ at $\theta=\pi$ is guaranteed (see supplemental section IV).

\emph{Experimental related issues.--} \label{sec:exp} 
The scaled disorder parameter $f(\theta)$ can be measured by counting $U(1)$ charges~\cite{sherson2010single} inside region $A$ for each sample and post-processing the statistical data. 
Specifically, after performing $N_M$ times projective measurements of the many-body state, we record the frequency distribution $q(n)$ of observing $n$ charges in $A$.
The disorder parameter is then estimated as
\begin{equation}
    \lrangle{U_A(\theta)} \approx \frac{1}{N_M}\sum_n q(n)\exp(\ii \theta  n)\,. 
\end{equation}

The protocol remains effective under noise, provided that the dominant noises preserve \emph{strong} $U(1)$ symmetry.
To see this, we model noise as a finite depth local unitary $V$ coupling the system to a environment state $\ket{0}_{\rm{E}}$: $\ket{\tilde{\psi}}=V(\ket{\psi}_{\rm{S}}\otimes \ket{0}_{\rm{E}})$.
Hence, $\ket{\tilde{\psi}}$ belongs to the same phase as $\ket{\psi}_{\rm{S}}$.
If $V$ commutes with $U(1)$ symmetry of the system -- forbidding charge exchange between the system and the environment -- then the measured charge statistics in $A$ remain uncontaminated, and our diagnosis is still valid (see supplemental section V for more details).

The projection onto the $n$-particle subspace $\hat{P}_n=\ket{n}\bra{n}$ suffers shot noise.
After $N_M$ measurements, the deviation of the empirical average $\mathbb{E}_{N_{M}}(\hat{P}_n)$ from the true probability $p_n$ is bounded by Chebyshev's inequality:
\begin{align}
    \mathrm{Pr}[|\mathbb{E}_{N_M}(\hat{P}_{n})-p_n|\geq \epsilon]
    \leq \frac{1}{4N_{M}\epsilon^2}\,.
    \label{eq:approx_prob}
\end{align}
Accurate resolving the non-analytic feature in $f(\theta)$ near $\theta=\pi$ requires controlling the statistical error $\epsilon\lesssim \ee^{-\abs{\partial A} f(\pi)}$.
Setting $\epsilon=\eta\ee^{-\abs{\partial A} f(\pi)}$, the number of measurements needed to achieve a failure probability below $\delta$ scales as $N_M\gtrsim \frac{1}{4\eta^2 \delta}e^{2f(\pi)\abs{\partial A}}$.
For example, in the bosonic correlated TI model investigate here, a $6\times6$ subregion with $\eta=\delta=0.1$ requires $N_M\sim 10^{7}$.
In a quantum processor with imaging time $\sim10^3\mu$s~\cite{bluvstein2024logical}, this corresponds to a few minutes of data acquisition.

Now, two key consideration arise: 
\begin{enumerate}
    \item The measurement cost to resolve $f(\theta)$ scales exponentially with $\abs{\partial A}$;
    \item The sub-leading corrections in Eq.~\eqref{eq:area_law_disorder_param} are suppressed only when $\abs{\partial A}\gg \xi$ , with $\xi$ the correlation length.
\end{enumerate}
These constraints restrict the protocol to intermediate region sizes -- a regime ideally matched to the capabilities of current noise intermediate-Scale quantum platforms~\cite{preskill2018quantum} and their associated algorithms~\cite{bharti2022noisy}.

\emph{Outlook.--}\label{sec:disc}
Our work reveals a fundamental duality between 2D correlated TI and 1D mixed state topological phases~\cite{ma2023c,xu2025diagnosing} realized by RDM at the entanglement boundary.
While we have focused on the $\ZZ_2$ Ising symmetric bosonic TI and the quantum spin Hall insulator, we anticipate this framework will lead to new diagnostic tools for a broader classes of correlated TI beyond the specific cases studied here, which we leave for future work.

A natural extension of this work involves generalizing this pure-mixed duality to broader classes of topological phases, such as symmetry-enriched topological~(SET) phases.
In such systems, the one-form symmetry of topological order manifests as the strong symmetry in the 1D mixed state, leading to interplay between strong and weak symmetries, potentially enabling the extraction of novel non-local order parameters of SET phases.

\emph{Acknowledgements.--}
We thank Biao Huang and Yingfei Gu for helpful discussions. 
This work is supported by MOST NO. 2022YFA1403902, NSFC Grant No.12574179. 
C.X. is also supported by the Shuimu Tsinghua Scholar program of Tsinghua University.

\emph{Note added.--}
Upon completing this work, we became aware of a related work by Shi, Sun, and Li~\cite{shi2025diagnose} that also studies the non-analytical behavior of the disorder parameter in 2D interacting topological insulators.

\bibliography{DisorderOperator}

@article{XuMoore2006,
  title = {Stability of the quantum spin Hall effect: Effects of interactions, disorder, and ${\mathbb{Z}}_{2}$ topology},
  author = {Xu, Cenke and Moore, J. E.},
  journal = {Phys. Rev. B},
  volume = {73},
  issue = {4},
  pages = {045322},
  numpages = {6},
  year = {2006},
  month = {Jan},
  publisher = {American Physical Society},
  doi = {10.1103/PhysRevB.73.045322},
  url = {https://link.aps.org/doi/10.1103/PhysRevB.73.045322}
}

@article{WuBernevigZhang2006,
  title = {Helical Liquid and the Edge of Quantum Spin Hall Systems},
  author = {Wu, Congjun and Bernevig, B. Andrei and Zhang, Shou-Cheng},
  journal = {Phys. Rev. Lett.},
  volume = {96},
  issue = {10},
  pages = {106401},
  numpages = {4},
  year = {2006},
  month = {Mar},
  publisher = {American Physical Society},
  doi = {10.1103/PhysRevLett.96.106401},
  url = {https://link.aps.org/doi/10.1103/PhysRevLett.96.106401}
}

@article{cai2024,
  title = {Universal contributions to charge fluctuations in spin chains at finite temperature},
  author = {Cai, Kang-Le and Cheng, Meng},
  journal = {Phys. Rev. B},
  volume = {111},
  issue = {20},
  pages = {205104},
  numpages = {19},
  year = {2025},
  month = {May},
  publisher = {American Physical Society},
  doi = {10.1103/PhysRevB.111.205104},
  url = {https://link.aps.org/doi/10.1103/PhysRevB.111.205104}
}

@article{chenSymmetryProtectedTopological2013,
  author = {Chen, Xie and Gu, Zheng-Cheng and Liu, Zheng-Xin and Wen, Xiao-Gang},
  title = {Symmetry Protected Topological Orders and the Group Cohomology of Their Symmetry Group},
  journal = {Phys. Rev. B},
  volume = {87},
  number = {15},
  pages = {155114},
  year = {2013},
  issn = {1098-0121, 1550-235X},
  doi = {10.1103/PhysRevB.87.155114},
  langid = {english}
}

@article{chenTopologicalDisorderParameter2022,
  author = {Chen, Bin-Bin and Tu, Hong-Hao and Meng, Zi Yang and Cheng, Meng},
  title = {Topological Disorder Parameter: A Many-Body Invariant to Characterize Gapped Quantum Phases},
  journal = {Phys. Rev. B},
  volume = {106},
  number = {9},
  pages = {094415},
  year = {2022},
  shorttitle = {Topological Disorder Parameter},
  publisher = {American Physical Society},
  doi = {10.1103/PhysRevB.106.094415}
}

@article{cian2021,
  author = {Cian, Ze-Pei and Dehghani, Hossein and Elben, Andreas and Vermersch, Benoît and Zhu, Guanyu},
  title = {Many-Body Chern Number from Statistical Correlations of Randomized Measurements},
  journal = {Phys. Rev. Lett.},
  volume = {126},
  number = {5},
  pages = {050501},
  year = {2021},
  issn = {0031-9007, 1079-7114},
  doi = {10.1103/PhysRevLett.126.050501},
  langid = {english}
}

@article{cirac2011,
  author = {Cirac, J. Ignacio and Poilblanc, Didier and Schuch, Norbert and Verstraete, Frank},
  title = {Entanglement Spectrum and Boundary Theories with Projected Entangled-Pair States},
  journal = {Phys. Rev. B},
  volume = {83},
  number = {24},
  pages = {245134},
  year = {2011},
  issn = {1098-0121, 1550-235X},
  doi = {10.1103/PhysRevB.83.245134}
}

@article{cai2024disorder,
  title = {Disorder operators in two-dimensional Fermi and non-Fermi liquids through multidimensional bosonization},
  author = {Cai, Kang-Le and Cheng, Meng},
  journal = {Phys. Rev. B},
  volume = {112},
  issue = {15},
  pages = {155123},
  numpages = {14},
  year = {2025},
  month = {Oct},
  publisher = {American Physical Society},
  doi = {10.1103/d8q8-4r7z},
  url = {https://link.aps.org/doi/10.1103/d8q8-4r7z}
}

@article{deleseleuc2019,
  author = {Léséleuc, Sylvain and Lienhard, Vincent and Scholl, Pascal and Barredo, Daniel and Weber, Sebastian},
  title = {Observation of a Symmetry-Protected Topological Phase of Interacting Bosons with Rydberg Atoms},
  journal = {Science},
  volume = {365},
  number = {6455},
  pages = {775--780},
  year = {2019},
  publisher = {American Association for the Advancement of Science},
  doi = {10.1126/science.aav9105}
}

@article{schafer2020tools,
  title={Tools for quantum simulation with ultracold atoms in optical lattices},
  author={Sch{\"a}fer, Florian and Fukuhara, Takeshi and Sugawa, Seiji and Takasu, Yosuke and Takahashi, Yoshiro},
  journal={Nature Reviews Physics},
  volume={2},
  number={8},
  pages={411--425},
  year={2020},
  publisher={Nature Publishing Group UK London},
  url={https://doi.org/10.1038/s42254-020-0195-3}
}

@article{ebadi2021,
  author = {Ebadi, Sepehr and Wang, Tout T. and Levine, Harry and Keesling, Alexander and Semeghini, Giulia},
  title = {Quantum Phases of Matter on a 256-Atom Programmable Quantum Simulator},
  journal = {Nature},
  volume = {595},
  number = {7866},
  pages = {227--232},
  year = {2021},
  issn = {1476-4687},
  publisher = {Nature Publishing Group},
  doi = {10.1038/s41586-021-03582-4},
  langid = {english}
}

@article{elben2020,
  author = {Elben, Andreas and Yu, Jinlong and Zhu, Guanyu and Hafezi, Mohammad and Pollmann, Frank and Zoller, Peter and Vermersch, Benoît},
  title = {Many-Body Topological Invariants from Randomized Measurements in Synthetic Quantum Matter},
  journal = {Sci. Adv.},
  volume = {6},
  number = {15},
  pages = {eaaz3666},
  year = {2020},
  issn = {2375-2548},
  doi = {10.1126/sciadv.aaz3666},
  langid = {english}
}

@article{preskill2018quantum,
  title={Quantum computing in the NISQ era and beyond},
  author={Preskill, John},
  journal={Quantum},
  volume={2},
  pages={79},
  year={2018},
  publisher={Verein zur F{\"o}rderung des Open Access Publizierens in den Quantenwissenschaften},
  url={https://doi.org/10.22331/q-2018-08-06-79}
}

@article{goldman2010,
  author = {Goldman, N. and Satija, I. and Nikolic, P. and Bermudez, A. and Martin-Delgado, M. A. and Lewenstein, M. and Spielman, I. B.},
  title = {Realistic Time-Reversal Invariant Topological Insulators with Neutral Atoms},
  journal = {Phys. Rev. Lett.},
  volume = {105},
  number = {25},
  pages = {255302},
  year = {2010},
  publisher = {American Physical Society},
  doi = {10.1103/PhysRevLett.105.255302}
}

@article{GuWen2009,
  title = {Tensor-entanglement-filtering renormalization approach and symmetry-protected topological order},
  author = {Gu, Zheng-Cheng and Wen, Xiao-Gang},
  journal = {Phys. Rev. B},
  volume = {80},
  issue = {15},
  pages = {155131},
  numpages = {23},
  year = {2009},
  month = {Oct},
  publisher = {American Physical Society},
  doi = {10.1103/PhysRevB.80.155131},
  url = {https://link.aps.org/doi/10.1103/PhysRevB.80.155131}
}

@article{goldman2016,
  author = {Goldman, N. and Budich, J. C. and Zoller, P.},
  title = {Topological Quantum Matter with Ultracold Gases in Optical Lattices},
  journal = {Nature Phys},
  volume = {12},
  number = {7},
  pages = {639--645},
  year = {2016},
  issn = {1745-2473, 1745-2481},
  doi = {10.1038/nphys3803}
}

@article{bluvstein2024logical,
  title={Logical quantum processor based on reconfigurable atom arrays},
  author={Bluvstein, Dolev and Evered, Simon J and Geim, Alexandra A and Li, Sophie H and Zhou, Hengyun and Manovitz, Tom and Ebadi, Sepehr and Cain, Madelyn and Kalinowski, Marcin and Hangleiter, Dominik and others},
  journal={Nature},
  volume={626},
  number={7997},
  pages={58--65},
  year={2024},
  publisher={Nature Publishing Group UK London},
  url={https://doi.org/10.1038/s41586-023-06927-3}
}

@article{semeghini2021probing,
  title={Probing topological spin liquids on a programmable quantum simulator},
  author={Semeghini, Giulia and Levine, Harry and Keesling, Alexander and Ebadi, Sepehr and Wang, Tout T and Bluvstein, Dolev and Verresen, Ruben and Pichler, Hannes and Kalinowski, Marcin and Samajdar, Rhine and others},
  journal={Science},
  volume={374},
  number={6572},
  pages={1242--1247},
  year={2021},
  publisher={American Association for the Advancement of Science},
  doi={10.1126/science.abi8794}
}

@article{iqbal2024topological,
  title={Topological order from measurements and feed-forward on a trapped ion quantum computer},
  author={Iqbal, Mohsin and Tantivasadakarn, Nathanan and Gatterman, Thomas M and Gerber, Justin A and Gilmore, Kevin and Gresh, Dan and Hankin, Aaron and Hewitt, Nathan and Horst, Chandler V and Matheny, Mitchell and others},
  journal={Communications Physics},
  volume={7},
  number={1},
  pages={205},
  year={2024},
  publisher={Nature Publishing Group UK London},
  doi={10.1038/s42005-024-01698-3}
}

@article{satzinger2021realizing,
  title={Realizing topologically ordered states on a quantum processor},
  author={Satzinger, KJ and Liu, Y-J and Smith, A and Knapp, C and Newman, M and Jones, C and Chen, Z and Quintana, C and Mi, X and Dunsworth, A and others},
  journal={Science},
  volume={374},
  number={6572},
  pages={1237--1241},
  year={2021},
  publisher={American Association for the Advancement of Science},
  doi={10.1126/science.abi8378},
}

@article{sullivan2020interacting,
  title={Interacting edge states of fermionic symmetry-protected topological phases in two dimensions},
  author={Sullivan, Joseph and Cheng, Meng},
  journal={SciPost Phys.},
  volume={9},
  number={2},
  pages={016},
  year={2020},
  doi={10.21468/SciPostPhys.9.2.016},
}

@article{ning2021edge,
  title={Edge theories of two-dimensional fermionic symmetry protected topological phases protected by unitary Abelian symmetries},
  author={Ning, Shang-Qiang and Wang, Chenjie and Wang, Qing-Rui and Gu, Zheng-Cheng},
  journal={Physical Review B},
  volume={104},
  number={7},
  pages={075151},
  year={2021},
  publisher={APS}
}

@misc{Metlitski20191d,
    title={A 1d lattice model for the boundary of the quantum spin-Hall insulator}, 
    author={Max A. Metlitski},
    year={2019},
    eprint={1908.08958},
    archivePrefix={arXiv},
    primaryClass={cond-mat.str-el}
}

@article{jiang2015,
  author = {Jiang, Shenghan and Ran, Ying},
  title = {Symmetric Tensor Networks and Practical Simulation Algorithms to Sharply Identify Classes of Quantum Phases Distinguishable by Short-Range Physics},
  journal = {Phys. Rev. B},
  volume = {92},
  number = {10},
  pages = {104414},
  year = {2015},
  issn = {1098-0121, 1550-235X},
  doi = {10.1103/PhysRevB.92.104414},
  langid = {english}
}

@article{jiang2023,
  author = {Jiang, Weilun and Chen, Bin-Bin and Liu, Zi Hong and Rong, Junchen and Assaad, Fakher F.},
  title = {Many versus One: The Disorder Operator and Entanglement Entropy in Fermionic Quantum Matter},
  journal = {SciPost Phys.},
  volume = {15},
  number = {3},
  pages = {082},
  year = {2023},
  issn = {2542-4653},
  shorttitle = {Many versus One},
  doi = {10.21468/SciPostPhys.15.3.082},
  langid = {english}
}

@article{jiangAnyonCondensationGeneric2017,
  author = {Jiang, Shenghan and Ran, Ying},
  title = {Anyon Condensation and a Generic Tensor-Network Construction for Symmetry Protected Topological Phases},
  journal = {Phys. Rev. B},
  volume = {95},
  number = {12},
  pages = {125107},
  year = {2017},
  issn = {2469-9950, 2469-9969},
  doi = {10.1103/PhysRevB.95.125107},
  langid = {english}
}

@article{levin2006detecting,
  title={Detecting topological order in a ground state wave function},
  author={Levin, Michael and Wen, Xiao-Gang},
  journal = {Phys. Rev. Lett.},
  volume = {96},
  issue = {11},
  pages = {110405},
  numpages = {4},
  year = {2006},
  month = {Mar},
  publisher = {American Physical Society},
  doi = {10.1103/PhysRevLett.96.110405},
  url = {https://link.aps.org/doi/10.1103/PhysRevLett.96.110405}
}

@article{kitaev2006,
  author = {Kitaev, Alexei and Preskill, John},
  title = {Topological Entanglement Entropy},
  journal = {Phys. Rev. Lett.},
  volume = {96},
  number = {11},
  pages = {110404},
  year = {2006},
  issn = {0031-9007, 1079-7114},
  doi = {10.1103/PhysRevLett.96.110404},
  langid = {english}
}

@article{levin2012,
  author = {Levin, Michael and Gu, Zheng-Cheng},
  title = {Braiding Statistics Approach to Symmetry-Protected Topological Phases},
  journal = {Phys. Rev. B},
  volume = {86},
  number = {11},
  pages = {115109},
  year = {2012},
  issn = {1098-0121, 1550-235X},
  doi = {10.1103/PhysRevB.86.115109},
  langid = {english}
}

@article{evenbly2015tensor,
  title={Tensor network renormalization},
  author={Evenbly, Glen and Vidal, Guifre},
  journal = {Phys. Rev. Lett.},
  volume = {115},
  issue = {18},
  pages = {180405},
  numpages = {6},
  year = {2015},
  month = {Oct},
  publisher = {American Physical Society},
  doi = {10.1103/PhysRevLett.115.180405},
  url = {https://link.aps.org/doi/10.1103/PhysRevLett.115.180405}
}

@article{iqbal2024non,
  title={Non-Abelian topological order and anyons on a trapped-ion processor},
  author={Iqbal, Mohsin and Tantivasadakarn, Nathanan and Verresen, Ruben and Campbell, Sara L and Dreiling, Joan M and Figgatt, Caroline and Gaebler, John P and Johansen, Jacob and Mills, Michael and Moses, Steven A and others},
  journal={Nature},
  volume={626},
  number={7999},
  pages={505--511},
  year={2024},
  publisher={Nature Publishing Group UK London},
  doi={10.1038/s41586-023-06934-4}
}

@article{liu2014,
  author = {Liu, Zheng-Xin and Gu, Zheng-Cheng and Wen, Xiao-Gang},
  title = {Microscopic Realization of 2-Dimensional Bosonic Topological Insulators},
  journal = {Phys. Rev. Lett.},
  volume = {113},
  number = {26},
  pages = {267206},
  year = {2014},
  issn = {0031-9007, 1079-7114},
  doi = {10.1103/PhysRevLett.113.267206}
}

@online{ma2023c,
  author = {Ma, Ruochen and Zhang, Jian-Hao and Bi, Zhen and Cheng, Meng and Wang, Chong},
  title = {Topological Phases with Average Symmetries: The Decohered, the Disordered, and the Intrinsic},
  number = {arXiv:2305.16399},
  year = {2023},
  url = {http://arxiv.org/abs/2305.16399},
  shorttitle = {Topological Phases with Average Symmetries},
  pubstate = {preprint}
}

@article{ma2024variational,
  title={Variational Tensor Wave Functions for the Interacting Quantum Spin Hall Phase},
  author={Ma, Yixin and Jiang, Shenghan and Xu, Chao},
  journal = {Phys. Rev. Lett.},
  volume = {132},
  issue = {12},
  pages = {126504},
  numpages = {7},
  year = {2024},
  month = {Mar},
  publisher = {American Physical Society},
  doi = {10.1103/PhysRevLett.132.126504},
  url = {https://link.aps.org/doi/10.1103/PhysRevLett.132.126504}
}

@article{pollmann2012,
  author = {Pollmann, Frank and Turner, Ari M.},
  title = {Detection of Symmetry Protected Topological Phases in 1D},
  journal = {Phys. Rev. B},
  volume = {86},
  number = {12},
  pages = {125441},
  year = {2012},
  issn = {1098-0121, 1550-235X},
  doi = {10.1103/PhysRevB.86.125441},
  langid = {english}
}

@article{wang2023exactly,
  title = {Exactly solvable lattice models for interacting electronic insulators in two dimensions},
  author = {Wang, Qing-Rui and Qi, Yang and Fang, Chen and Cheng, Meng and Gu, Zheng-Cheng},
  journal = {Phys. Rev. B},
  volume = {108},
  issue = {12},
  pages = {L121104},
  numpages = {6},
  year = {2023},
  month = {Sep},
  publisher = {American Physical Society},
  doi = {10.1103/PhysRevB.108.L121104},
  url = {https://link.aps.org/doi/10.1103/PhysRevB.108.L121104}
}

@article{ChenGuWen2011,
    title = {Classification of gapped symmetric phases in one-dimensional spin systems},
    author = {Chen, Xie and Gu, Zheng-Cheng and Wen, Xiao-Gang},
    journal = {Phys. Rev. B},
    volume = {83},
    issue = {3},
    pages = {035107},
    numpages = {19},
    year = {2011},
    month = {Jan},
    publisher = {American Physical Society},
    doi = {10.1103/PhysRevB.83.035107},
    url = {https://link.aps.org/doi/10.1103/PhysRevB.83.035107}
}

@article{bernevig2006quantum,
  title={Quantum spin Hall effect and topological phase transition in HgTe quantum wells},
  author={Bernevig, B Andrei and Hughes, Taylor L and Zhang, Shou-Cheng},
  journal={Science},
  volume={314},
  number={5806},
  pages={1757--1761},
  year={2006},
  publisher={American Association for the Advancement of Science},
  doi={10.1126/science.1133734}
}

@article{sherson2010single,
  title={Single-atom-resolved fluorescence imaging of an atomic Mott insulator},
  author={Sherson, Jacob F and Weitenberg, Christof and Endres, Manuel and Cheneau, Marc and Bloch, Immanuel and Kuhr, Stefan},
  journal={Nature},
  volume={467},
  number={7311},
  pages={68--72},
  year={2010},
  publisher={Nature Publishing Group UK London},
 url={https://doi.org/10.1038/nature09378}
}

@article{KaneMele2005,
  title = {Quantum Spin Hall Effect in Graphene},
  author = {Kane, C. L. and Mele, E. J.},
  journal = {Phys. Rev. Lett.},
  volume = {95},
  issue = {22},
  pages = {226801},
  numpages = {4},
  year = {2005},
  month = {Nov},
  publisher = {American Physical Society},
  doi = {10.1103/PhysRevLett.95.226801},
  url = {https://link.aps.org/doi/10.1103/PhysRevLett.95.226801}
}

@article{QiZhang2011,
  title = {Topological insulators and superconductors},
  author = {Qi, Xiao-Liang and Zhang, Shou-Cheng},
  journal = {Rev. Mod. Phys.},
  volume = {83},
  issue = {4},
  pages = {1057--1110},
  numpages = {0},
  year = {2011},
  month = {Oct},
  publisher = {American Physical Society},
  doi = {10.1103/RevModPhys.83.1057},
  url = {https://link.aps.org/doi/10.1103/RevModPhys.83.1057}
}

@article{HasanKane2010,
  title = {Colloquium: Topological insulators},
  author = {Hasan, M. Z. and Kane, C. L.},
  journal = {Rev. Mod. Phys.},
  volume = {82},
  issue = {4},
  pages = {3045--3067},
  numpages = {0},
  year = {2010},
  month = {Nov},
  publisher = {American Physical Society},
  doi = {10.1103/RevModPhys.82.3045},
  url = {https://link.aps.org/doi/10.1103/RevModPhys.82.3045}
}

@article{ChenGuLiuWen2012Symmetry,
  title={Symmetry-protected topological orders in interacting bosonic systems},
  author={Chen, Xie and Gu, Zheng-Cheng and Liu, Zheng-Xin and Wen, Xiao-Gang},
  journal={Science},
  volume={338},
  number={6114},
  pages={1604--1606},
  year={2012},
  doi = {10.1126/science.1227224},
  publisher={American Association for the Advancement of Science}
}

@article{senthil2015symmetry,
  title={Symmetry-protected topological phases of quantum matter},
  author={Senthil, Todadri},
  journal={Annu. Rev. Condens. Matter Phys.},
  volume={6},
  number={1},
  pages={299--324},
  year={2015},
  publisher={Annual Reviews},
  url = {https://www.annualreviews.org/doi/abs/10.1146/annurev-conmatphys-031214-014740}
}

@article{tang2017quantum,
  title={Quantum spin Hall state in monolayer 1T'-WTe2},
  author={Tang, Shujie and Zhang, Chaofan and Wong, Dillon and Pedramrazi, Zahra and Tsai, Hsin-Zon and Jia, Chunjing and Moritz, Brian and Claassen, Martin and Ryu, Hyejin and Kahn, Salman and others},
  journal={Nature Physics},
  volume={13},
  number={7},
  pages={683--687},
  year={2017},
  publisher={Nature Publishing Group UK London},
  url={https://doi.org/10.1038/nphys4174}
}

@article{yao2023observation,
  title={Observation of many-body Fock space dynamics in two dimensions},
  author={Yao, Yunyan and Xiang, Liang and Guo, Zexian and Bao, Zehang and Yang, Yong-Feng and Song, Zixuan and Shi, Haohai and Zhu, Xuhao and Jin, Feitong and Chen, Jiachen and others},
  journal={Nature Physics},
  volume={19},
  number={10},
  pages={1459--1465},
  year={2023},
  publisher={Nature Publishing Group UK London}
}

@article{fei2017edge,
  title={Edge conduction in monolayer WTe2},
  author={Fei, Zaiyao and Palomaki, Tauno and Wu, Sanfeng and Zhao, Wenjin and Cai, Xinghan and Sun, Bosong and Nguyen, Paul and Finney, Joseph and Xu, Xiaodong and Cobden, David H},
  journal={Nature Physics},
  volume={13},
  number={7},
  pages={677--682},
  year={2017},
  publisher={Nature Publishing Group UK London},
  url={https://doi.org/10.1038/nphys4091}
}

@article{perez2008string,
  title={String order and symmetries in quantum spin lattices},
  author={P{\'e}rez-Garc{\'\i}a, David and Wolf, Michael M and Sanz, M and Verstraete, Frank and Cirac, J Ignacio},
  journal={Phys. Rev. Lett.},
  volume={100},
  number={16},
  pages={167202},
  year={2008},
  publisher={APS},
  url={https://doi.org/10.1103/PhysRevLett.100.167202}
}

@article{sompet2022a,
  author = {Sompet, Pimonpan and Hirthe, Sarah and Bourgund, Dominik and Chalopin, Thomas and Bibo, Julian},
  title = {Realizing the Symmetry-Protected Haldane Phase in Fermi–Hubbard Ladders},
  journal = {Nature},
  volume = {606},
  number = {7914},
  pages = {484--488},
  year = {2022},
  issn = {1476-4687},
  publisher = {Nature Publishing Group},
  doi = {10.1038/s41586-022-04688-z},
  langid = {english}
}

@article{wangBosonicAnomaliesInduced2015,
  author = {Wang, Juven and Santos, Luiz H. and Wen, Xiao-Gang},
  title = {Bosonic Anomalies, Induced Fractional Quantum Numbers and Degenerate Zero Modes: The Anomalous Edge Physics of Symmetry-Protected Topological States},
  journal = {Phys. Rev. B},
  volume = {91},
  number = {19},
  pages = {195134},
  year = {2015},
  issn = {1098-0121, 1550-235X},
  shorttitle = {Bosonic Anomalies, Induced Fractional Quantum Numbers and Degenerate Zero Modes},
  doi = {10.1103/PhysRevB.91.195134}
}

@article{wen2017,
  author = {Wen, Xiao-Gang},
  title = {Zoo of Quantum-Topological Phases of Matter},
  journal = {Rev. Mod. Phys.},
  volume = {89},
  number = {4},
  pages = {041004},
  year = {2017},
  issn = {0034-6861, 1539-0756},
  doi = {10.1103/RevModPhys.89.041004}
}

@article{xu2024unveiling,
  title={Unveiling correlated two-dimensional topological insulators through fermionic tensor network states—classification, edge theories and variational wavefunctions},
  author={Xu, Chao and Ma, Yixin and Jiang, Shenghan},
  journal={Reports on Progress in Physics},
  volume={87},
  number={10},
  pages={108001},
  year={2024},
  publisher={IOP Publishing},
  url={https://iopscience.iop.org/article/10.1088/1361-6633/ad7058}
}

@article{chang2013experimental,
  title={Experimental observation of the quantum anomalous Hall effect in a magnetic topological insulator},
  author={Chang, Cui-Zu and Zhang, Jinsong and Feng, Xiao and Shen, Jie and Zhang, Zuocheng and Guo, Minghua and Li, Kang and Ou, Yunbo and Wei, Pang and Wang, Li-Li and others},
  journal={Science},
  volume={340},
  number={6129},
  pages={167--170},
  year={2013},
  publisher={American Association for the Advancement of Science},
 url={10.1126/science.1234414}
}

@article{pichler2016measure,
  title = {Measurement Protocol for the Entanglement Spectrum of Cold Atoms},
  author = {Pichler, Hannes and Zhu, Guanyu and Seif, Alireza and Zoller, Peter and Hafezi, Mohammad},
  journal = {Phys. Rev. X},
  volume = {6},
  issue = {4},
  pages = {041033},
  numpages = {12},
  year = {2016},
  month = {Nov},
  publisher = {American Physical Society},
  doi = {10.1103/PhysRevX.6.041033},
  url = {https://link.aps.org/doi/10.1103/PhysRevX.6.041033}
}

@article{andersen2019entanglement,
  title={Entanglement stabilization using ancilla-based parity detection and real-time feedback in superconducting circuits},
  author={Andersen, Christian Kraglund and Remm, Ants and Lazar, Stefania and Krinner, Sebastian and Heinsoo, Johannes and Besse, Jean-Claude and Gabureac, Mihai and Wallraff, Andreas and Eichler, Christopher},
  journal={npj Quantum Information},
  volume={5},
  number={1},
  pages={69},
  year={2019},
  publisher={Nature Publishing Group UK London},
  url={https://doi.org/10.1038/s41534-019-0185-4}
}

@article{Grusdt2016,
	author = {Grusdt, F. and Yao, N. Y. and Abanin, D. and Fleischhauer, M. and Demler, E.},
	date = {2016/06/17},
	doi = {10.1038/ncomms11994},
	id = {Grusdt2016},
	isbn = {2041-1723},
	journal = {Nature Communications},
	number = {1},
	pages = {11994},
	title = {Interferometric measurements of many-body topological invariants using mobile impurities},
	url = {https://doi.org/10.1038/ncomms11994},
	volume = {7},
	year = {2016}
}

@article{bharti2022noisy,
  title = {Noisy intermediate-scale quantum algorithms},
  author = {Bharti, Kishor and Cervera-Lierta, Alba and Kyaw, Thi Ha and Haug, Tobias and Alperin-Lea, Sumner and Anand, Abhinav and Degroote, Matthias and Heimonen, Hermanni and Kottmann, Jakob S. and Menke, Tim and Mok, Wai-Keong and Sim, Sukin and Kwek, Leong-Chuan and Aspuru-Guzik, Al\'an},
  journal = {Rev. Mod. Phys.},
  volume = {94},
  issue = {1},
  pages = {015004},
  numpages = {69},
  year = {2022},
  month = {Feb},
  publisher = {American Physical Society},
  doi = {10.1103/RevModPhys.94.015004},
  url = {https://link.aps.org/doi/10.1103/RevModPhys.94.015004}
}

@article{clarke2008superconducting,
  title={Superconducting quantum bits},
  author={Clarke, John and Wilhelm, Frank K},
  journal={Nature},
  volume={453},
  number={7198},
  pages={1031--1042},
  year={2008},
  publisher={Nature Publishing Group UK London},
  url={https://doi.org/10.1038/nature07128}
}

@article{reagor2018demonstration,
  title={Demonstration of universal parametric entangling gates on a multi-qubit lattice},
  author={Reagor, Matthew and Osborn, Christopher B and Tezak, Nikolas and Staley, Alexa and Prawiroatmodjo, Guenevere and Scheer, Michael and Alidoust, Nasser and Sete, Eyob A and Didier, Nicolas and da Silva, Marcus P and others},
  journal={Science advances},
  volume={4},
  number={2},
  pages={eaao3603},
  year={2018},
  publisher={American Association for the Advancement of Science},
  url={10.1126/sciadv.aao3603}
}

@article{debnath2016demonstration,
  title={Demonstration of a small programmable quantum computer with atomic qubits},
  author={Debnath, Shantanu and Linke, Norbert M and Figgatt, Caroline and Landsman, Kevin A and Wright, Kevin and Monroe, Christopher},
  journal={Nature},
  volume={536},
  number={7614},
  pages={63--66},
  year={2016},
  publisher={Nature Publishing Group UK London},
  url={https://doi.org/10.1038/nature18648}
}

@article{williamson2016matrix,
  title={Matrix product operators for symmetry-protected topological phases: Gauging and edge theories},
  author={Williamson, Dominic J and Bultinck, Nick and Mari{\"e}n, Michael and {\c{S}}ahino{\u{g}}lu, Mehmet B and Haegeman, Jutho and Verstraete, Frank},
  journal={Physical Review B},
  volume={94},
  number={20},
  pages={205150},
  year={2016},
  publisher={APS},
  url={https://doi.org/10.1103/PhysRevB.94.205150}
}

@article{chen2011two,
  title = {Two-dimensional symmetry-protected topological orders and their protected gapless edge excitations},
  author = {Chen, Xie and Liu, Zheng-Xin and Wen, Xiao-Gang},
  journal = {Phys. Rev. B},
  volume = {84},
  issue = {23},
  pages = {235141},
  numpages = {13},
  year = {2011},
  month = {Dec},
  publisher = {American Physical Society},
  doi = {10.1103/PhysRevB.84.235141},
  url = {https://link.aps.org/doi/10.1103/PhysRevB.84.235141}
}

@article{friis2018observation,
  title={Observation of entangled states of a fully controlled 20-qubit system},
  author={Friis, Nicolai and Marty, Oliver and Maier, Christine and Hempel, Cornelius and Holz{\"a}pfel, Milan and Jurcevic, Petar and Plenio, Martin B and Huber, Marcus and Roos, Christian and Blatt, Rainer and others},
  journal={Physical Review X},
  volume={8},
  number={2},
  pages={021012},
  year={2018},
  publisher={APS},
  url={https://doi.org/10.1103/PhysRevX.8.021012}
}

@article{zhang2017observation,
  title={Observation of a many-body dynamical phase transition with a 53-qubit quantum simulator},
  author={Zhang, Jiehang and Pagano, Guido and Hess, Paul W and Kyprianidis, Antonis and Becker, Patrick and Kaplan, Harvey and Gorshkov, Alexey V and Gong, Z-X and Monroe, Christopher},
  journal={Nature},
  volume={551},
  number={7682},
  pages={601--604},
  year={2017},
  publisher={Nature Publishing Group UK London},
  url={https://doi.org/10.1038/nature24654}
}

@article{wright2019benchmarking,
  title={Benchmarking an 11-qubit quantum computer},
  author={Wright, Kenneth and Beck, Kristin M and Debnath, Sea and Amini, JM and Nam, Y and Grzesiak, N and Chen, J-S and Pisenti, NC and Chmielewski, M and Collins, C and others},
  journal={Nature communications},
  volume={10},
  number={1},
  pages={5464},
  year={2019},
  publisher={Nature Publishing Group UK London},
  url={https://doi.org/10.1038/s41467-019-13534-2}
}

@misc{SM,
  title = {See {S}upplemental {M}aterials for more details on the symmetries of the RDM, PEPS representation of the SPT wavefunctions, MPDO representaions of the RDM, the corresponding tensor equations for cTIs, and the noise analysis.}
}

@article{xu2025diagnosing,
  title={Diagnosing 2D symmetry protected topological states via mixed state anomaly},
  author={Xu, Chao and Zang, Yunlong and Ma, Yixin and Gu, Yingfei and Jiang, Shenghan},
  journal={arXiv preprint arXiv:2506.13096},
  year={2025},
  url={https://doi.org/10.48550/arXiv.2506.13096
}
}

@article{wang2024distinguishing,
  title={Distinguishing quantum phases through cusps in full counting statistics},
  author={Wang, Chang-Yan and Zhou, Tian-Gang and Zhou, Yi-Neng and Zhang, Pengfei},
  journal={Physical Review Letters},
  volume={133},
  number={8},
  pages={083402},
  year={2024},
  publisher={APS},
  url={https://doi.org/10.1103/PhysRevLett.133.083402}
}

@article{wen2023bulk,
  title={Bulk-boundary correspondence for intrinsically gapless symmetry-protected topological phases from group cohomology},
  author={Wen, Rui and Potter, Andrew C},
  journal={Physical Review B},
  volume={107},
  number={24},
  pages={245127},
  year={2023},
  publisher={APS},
  url={https://doi.org/10.1103/PhysRevB.107.245127}
}

@article{mao2025probing,
  title={Probing the Topology of Fermionic Gaussian Mixed States with U (1) symmetry by Full Counting Statistics},
  author={Mao, Liang and Zhai, Hui and Yang, Fan},
  journal={Chinese Physics Letters},
  volume={42},
  number={6},
  pages={067401},
  year={2025},
  publisher={IOP Publishing},
  url={https://iopscience.iop.org/article/10.1088/0256-307X/42/6/067401}
}

@misc{horinouchi2020solvablelatticemodel21d,
      title={Solvable lattice model for (2+1)D bosonic topological insulator}, 
      author={Yusuke Horinouchi},
      year={2020},
      eprint={2002.01639},
      archivePrefix={arXiv},
      primaryClass={cond-mat.mes-hall},
      url={https://arxiv.org/abs/2002.01639}, 
}

@article{wang2022exactlyu1set,
  title = {Exactly solvable models for U(1) symmetry-enriched topological phases},
  author = {Wang, Qing-Rui and Cheng, Meng},
  journal = {Phys. Rev. B},
  volume = {106},
  issue = {11},
  pages = {115104},
  numpages = {13},
  year = {2022},
  month = {Sep},
  publisher = {American Physical Society},
  doi = {10.1103/PhysRevB.106.115104},
  url = {https://link.aps.org/doi/10.1103/PhysRevB.106.115104}
}

@article{schuch2010peps,
  title={PEPS as ground states: Degeneracy and topology},
  author={Schuch, Norbert and Cirac, Ignacio and P{\'e}rez-Garc{\'\i}a, David},
  journal={Annals of Physics},
  volume={325},
  number={10},
  pages={2153--2192},
  year={2010},
  publisher={Elsevier},
  url={
https://doi.org/10.1016/j.aop.2010.05.008
}
}

@misc{shi2025diagnose,
  title = {Diagnose of mixed-state topological phases in strongly correlated systems by disorder parameter, to appear},
  author = {Shi, Shao-Hang and Sun, Xiao-Qi and Li, Zi-Xiang},
}
\end{document}


\title{Measurement protocol for detecting correlated topological insulators in synthetic quantum systems}
\author{Yixin Ma}
\affiliation{Kavli Institute for Theoretical Sciences, University of Chinese Academy of Sciences, Beijing 100190, China}

\author{Chao Xu}
\affiliation{Institute for Advanced Study, Tsinghua University, Beijing 100084, China}

\author{Shenghan Jiang}
\affiliation{Kavli Institute for Theoretical Sciences, University of Chinese Academy of Sciences, Beijing 100190, China}

\date{\today}
\title{Supplemental Materials}

\maketitle

\onecolumngrid
\tableofcontents
\vspace{0.6cm}
In this supplemental material, we present the representation of symmetric reduced density matrix~(RDM) on entanglement legs in Sec.~\ref{app:rdm_ec}.
Sec.~\ref{app:peps_ti} details the projected entangled pair states (PEPS) framework for correlated topological insulators (correlated TI), which gives the anomalous symmetry action on the boundary.
Sec.~\ref{app:bcTI} analyzes the symmetries of the matrix product density operator~(MPDO) and disorder parameter evaluation for for bosonic correlated TI, while Sec.~\ref{app:fcTI} presents the fermionic case.
Finally, Sec.~\ref{app:noise} discusses the robustness of the measurement protocol against noises.

\section{Reduced density matrix on entanglement legs}\label{app:rdm_ec}
Here, we show that RDM of subregion $A$~($\rho_{A}$) is isometric to a density operator $\rho_{\partial A}$ in the entanglement space.

Consider a wavefunction $\ket{\psi}$ defined on Hilbert space $\HH_A\otimes\HH_{\bar{A}}$ with Schmidt decomposition $\ket{\psi}=\sum^{\chi_{A}}_{\alpha=1}\ee^{-\lambda_{\alpha}/2}\ket{\phi^{\alpha}}_A\ket{\phi^{\alpha}}_{\bar{A}}$ with $\{\ket{\phi}_{A/\bar{A}}\}$ being orthonormal.
Here, $\{\lambda_\alpha\}$ is the entanglement spectrum, and $\chi_A$ is the Schmidt rank.
By introducing the entanglement space $\mathbb{H}_{\partial A}$ -- a $\chi_A$ dimensional Hilbert space with basis $\{\ket{\xi^\alpha}_{\partial A}|\alpha=1,\ldots,\chi_A\}$ --, the Schmit decomposition of $\ket{\psi}$ now admits a tensor representation
\begin{equation}
\begin{aligned}
    \ket{\psi}&=\sum_{\alpha} (|\phi^\alpha\rag_A\lag\xi^\alpha|_{\partial A})\cdot (e^{-\lambda_\alpha/2} |\xi^\alpha\rag_{\partial A}|\xi^\alpha\rag_{\partial A}) \cdot (\lag\xi^{\alpha}|_{\partial A} |\varphi^\alpha\rag_{\bar{A}})\nonumber\\
    &=\Tr_{\partial A}[T_A\otimes \Lambda_{\partial A}\otimes T_{\bar{A}}]\,
    =\begin{tikzpicture}
    [scale=0.8,baseline=-4pt]
        \draw[thick] (-0.1,0.18)-- (-0.1,0.6);
        \node at (0,-0.15) {$T_{A}$};
        \draw[thick,cyan] (0.3,-0.15)--(0.8,-0.15);
        \node at (1.3,-0.15) {$\Lambda_{\partial A}$};
        \draw[thick,cyan] (1.8,-0.15) -- (2.4,-0.15);
        \node at (2.8,-0.15) {$T_{\bar{A}}$};
        \draw[thick] (2.7,0.18)-- (2.7,0.6);
    \end{tikzpicture}
\end{aligned}
    \label{eq:Schmidt}
\end{equation}
where $\Tr_{\partial A}$ denotes contracting the entanglement space $\HH_{\partial A}$.
$T_A$~($T_{\bar{A}}$) is an \emph{isometry map} from $\HH_{\partial A}$ to $\HH_{A}$~($\HH_{\bar{A}}$).
For the case where $\ket{\psi}$ is represented by PEPS, $\HH_{\partial A}$ is identified as the boundary virtual legs of $A$, i.e. $\HH_{\partial A}=\otimes_{j\in\partial A} \mathcal{H}_j$.

By contracting $\HH_{\bar{A}}$ of $\ket{\psi}\bra{\psi}$, and using the isometry property for $T_{\bar{A}}$, we obtain
\begin{equation}
\begin{aligned}
    \rho_A&=\Tr_{\bar{A}}|\psi\rag \lag\psi|
    =\Tr_{\partial A,\partial A^{\dagger}}\left( T_A\otimes \Lambda_{\partial A}\otimes \Lambda^{\dagger}_{\partial A} \otimes T^{\dagger}_{A} \right)\\
    &=\begin{tikzpicture}[scale=0.8,baseline=-4pt]
        \draw[thick] (-0.1,0.18)-- (-0.1,0.6);
        \node at (0,-0.15) {$T_{A}$};
        \draw[thick,cyan] (0.3,-0.15)--(0.8,-0.15);
        \node at (1.3,-0.15) {$\rho_{\partial A}$};
        \draw[thick,cyan] (1.8,-0.15) -- (2.4,-0.15);
        \node at (2.9,-0.15) {$T_{A}^\dg$};
        \draw[thick] (2.8,0.18)-- (2.8,0.6);
    \end{tikzpicture}
    \end{aligned}
    \label{eq:rhoA_def}
\end{equation}
where $\rho_{\partial A}\equiv\Lambda_{\partial A} \cdot \Lambda^{\dagger}_{\partial A}=\sum_{\alpha} \ee^{-\lambda_\alpha}\ket{\xi^\alpha}\bra{\xi^{\alpha}}$ is an density operator in the entanglement space, and is related to $\rho_{A}$ by the isometry $T_A$~\cite{cirac2011}.

Consider the case where $\ket{\psi}$ hosts symmetry $U(g)\mathcal{K}^{s(g)}=\left( U_A(g)\otimes U_{\bar{A}}(g) \right) \mathcal{K}^{s(g)}$ with $\KK$ complex conjugation and $s(g)=0/1$ for the unitary/anti-unitary operation.
Such symmetry action induces a gauge transformation $W_{\partial A}(g)$ in the entanglement space, leading to the constraints on tensors:
\begin{equation}
    U_{A/\bar{A}}(g)\mathcal{K}^{s(g)}\cdot  T_{A/\bar{A}}=T_{A/\bar{A}}\cdot W_{\partial A}(g)~,\quad
    (W_{\partial A}\otimes W_{\partial A}) \cdot \Lambda_{\partial A}={}^{\mathcal{K}^{s(g)}}\!\Lambda_{\partial A}
\end{equation}
where ${}^{\mathcal{K}^{s(g)}}\!\Lambda_{\partial A}\equiv \mathcal{K}^{s(g)}\Lambda_{\partial A}\mathcal{K}^{s(g)}$.
Therefore, $\rho_{\partial A}$ hosts the following weak symmetry 
\begin{align}
    W_{\partial A}(g)\cdot {}^{\mathcal{K}^{s(g)}}\!\rho_{\partial A} \cdot W^{\dagger}_{\partial A}(g)=\rho_{\partial A}
    \label{eq:entangle_weak_sym}
\end{align}

The above derivation can be generalized to the fermionic case.
Local fermionic Hilbert space are a $\ZZ_2$-graded Hilbert space equipped with a parity operator $\hat{F}\equiv\sum_j(-)^{\abs{j}}\ket{j}\bra{j}$, where $\abs{j}=0(1)$ for parity even(odd) state.
The many-body fermionic Hilbert space is composed through the fermionic tensor product $\otimes_{f}$, which satisfies
\begin{equation}
    \ket{i}\otimes_f\ket{j}\otimes_f\cdots =(-)^{\abs{i}\abs{j}}\ket{i}\otimes_f\ket{j}\otimes_f\cdots~,\quad
    \ket{i}\otimes_f\bra{j}\otimes_f\cdots =(-)^{\abs{i}\abs{j}}\bra{i}\otimes_f\ket{j}\otimes\cdots~,\quad
    \cdots\cdots
    \label{}
\end{equation}
The tensor contraction is performed through the fermionic partial trace $\fTr$, where 
\begin{equation}
    \fTr[\bra{i}\otimes_f\ket{j}]=(-)^{\abs{i}\abs{j}}\fTr[\ket{j}\otimes_f\bra{i}]=\delta_{ij}  \, ,
\end{equation}
In the present study, we focus on fermionic tensor networks with {\it all local tensors being parity even}, and hence contraction of local tensors give a parity even wavefunction, which is independent of the order of local tensors.

Therefore, the above derivation for the bosonic tensor networks can be readily adapted to the fermionic cases by replacing $\otimes$ and $\Tr$ with their fermionic counterparts, $\otimes_f$ and $\fTr$\cite{xu2024unveiling,ma2024variational}.
Consequently, Eq.~\eqref{eq:rhoA_def} and \eqref{eq:entangle_weak_sym} continue to hold for the fermionic case as well.

\section{PEPS representation and tensor equations of correlated TI phases}\label{app:peps_ti}
In this part, we briefly review the PEPS representation of correlated TI for both bosonic and fermionic system, which helps to derive the anomalous boundary symmetries of these topological phases.
We refer Ref.~\cite{jiang2015,ma2024variational,xu2025diagnosing} for more details.

\subsection{Tensor equations for bosonic correlated TI}
\begin{figure}
    \centering
    \includegraphics[width=1\linewidth]{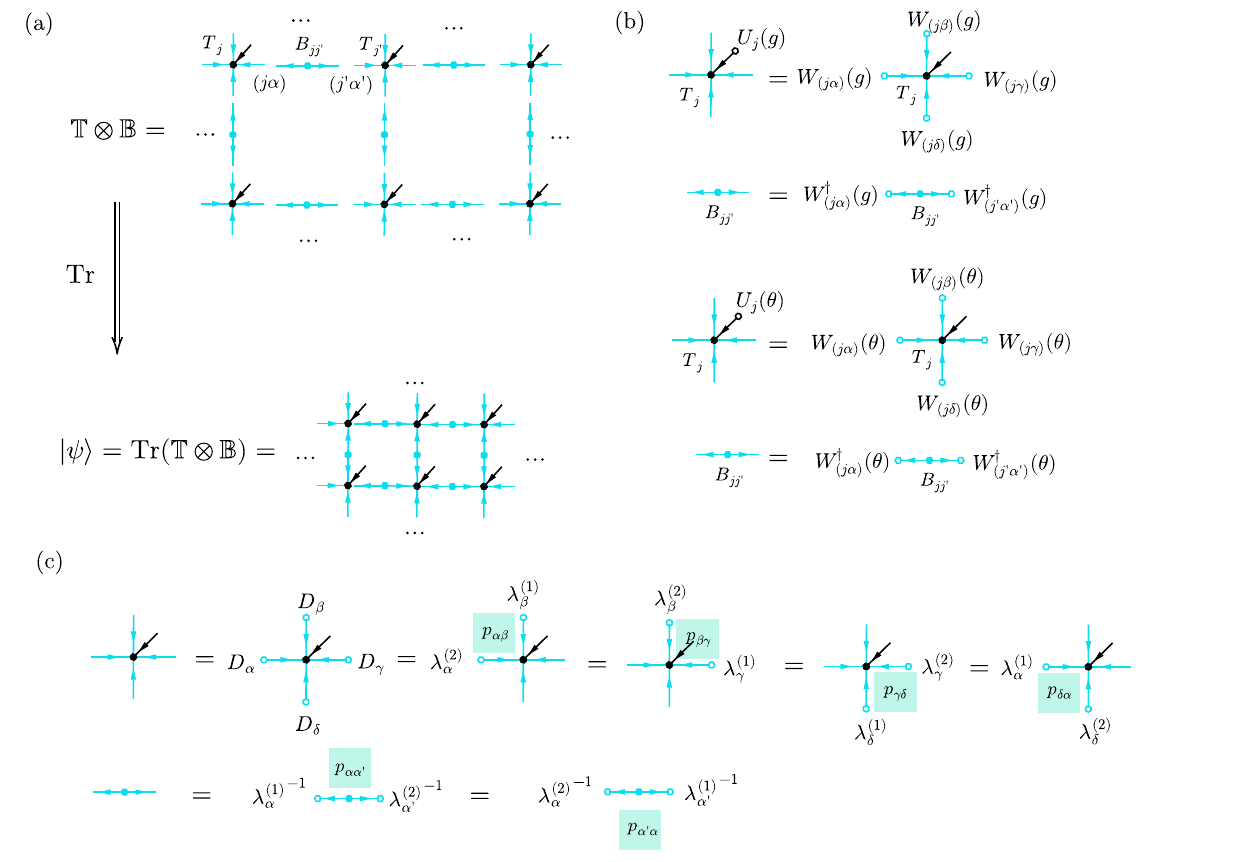}
    \caption{Symmetric PEPS for correlated TI. (a) A PEPS $|\psi\rag$ is constructed by contracting all site tensors $\mathbb{T}$ and all bond tensors $\mathbb{B}$.
    (b) The on-site unitary symmetries on physical legs induce gauge transformations on internal legs.
    (c) $\igg$ condition for local tensors of correlated TI.
    }
    \label{fig:tensor_eq}
\end{figure}

As shown in Fig.~\ref{fig:tensor_eq}(a), a PEPS state $\ket{\psi}$ is constructed with site tensors $\mathbb{T}=\otimes_j {T}_{j}$ and bond tensors $\mathbb{B}=\otimes_{\lrangle{jj'}} {B}_{jj'}$ where $i/j$ refers to lattice sites.
For the case where $\ket{\psi}$ is a bosonic insulator protected by the $SG=G \times U(1)$, the local tensors satisfy the following symmetric constraints
\begin{equation}
    U_j(g)\otimes\left( \bigotimes_{l=\alpha,\beta,\gamma,\delta}W_{(jl)}(g)\right) \KK^{s(g)}\cdot {T}_{j}= {T}_{j},\quad 
    {B}_{jj'} ={B}_{jj'}\cdot \KK^{s(g)}\left( W_{(j\alpha)}^{\dagger}(g)\otimes W^{\dagger}_{(j'\alpha')}(g)\right)~,\quad 
    \forall g\in G  
    \label{eq:sym_G}
\end{equation}
as well as
\begin{equation}
    U_j(\theta)\cdot {T}_{j} = \left(\bigotimes_{l}W_{(jl)}(\theta)\right) \cdot {T}_{j} ~,\quad 
    {B}_{jj'} ={B}_{jj'}\cdot \left( W_{(j\alpha)}^{\dagger}(\theta)\otimes W^{\dagger}_{(j'\alpha')}(\theta) \right)
\end{equation}
These constraints are depicted in Fig.~\ref{fig:tensor_eq}(b). 

Let $U_j(\theta)\equiv \ee ^{\ii\theta n_j}$ and $W_{(j\alpha)}(\theta)\equiv\ee^{\ii \theta n_{(j\alpha)}}$, we get
\begin{align}
    \left( n_j+\sum_l n_{(jl)} \right)\cdot T_j=0\,,\quad 
    B_{jj'}\cdot \left( n_{(j\alpha)}+n_{(j'\alpha')} \right) =0\,.
    \label{eq:sym_U}
\end{align}

Topological features of a state manifests as additional symmetries acting solely on internal legs \cite{schuch2010peps,jiang2015}, which are denoted as the invariant gauge group~($\igg$).
For example, for a topologically ordered state $\ket{\Phi}$, e.g. the toric code ground state, its local tensors holds virtual-leg symmetry $D$:
\begin{align}
    \left( \bigotimes_l D_{jl} \right) \cdot T_j=T_j~,\quad 
    B_{jj'}\cdot \left( D_{j\alpha} \otimes D_{j'\beta}^{\dg} \right) =B_{jj'}
\end{align}
where $D$ corresponds to the gauge flux loop.

To represent correlated TI, we introduce ``local plaquette" $\igg$, and ``global'' $\igg$ $D$ equals the product of local $\igg$ $\lambda$.
Physically, the decomposition to local $\igg$ is interpreted as condensing flux ~\cite{jiangAnyonCondensationGeneric2017}.
As shown in Fig.~\ref{fig:tensor_eq}(c), the local $\igg$ $\lambda$ imposes new constraints on local tensors.

The bosonic SPT protected by onsite unitary symmetry $SG$ is characterized by the cohomology class $[\omega]\in \mathcal{H}^{3}(SG,U(1))$~\cite{ChenGuLiuWen2012Symmetry}, which could be read from the consistent tensor equations.
To see this, we consider fusion of symmetry flux on virtual legs
\begin{align}
    W(g_1)W(g_2)=D(g_1,g_2)W(g_1g_2),\quad \forall g\in SG
\end{align}
where $D$ belongs to $\igg$, and we omit the leg index for simplicity.
The associativity of $W(g)$ leads to 
\begin{align}
    D(g_1,g_2)D(g_1g_2,g_3)=D(g_2,g_3)D(g_1,g_2g_3)
    \label{eq:igg_two_cocycle}
\end{align}
To impose the short-range entanglement condition, we require that $D_{\alpha}(g_1,g_2)$ could be decomposed to local $\igg$ $\lambda^{(1)}(g_1,g_2)\lambda^{(2)}(g_1,g_2)$.
Note that such decomposition is defined up to a global $U(1)$ phase factor. 
Therefore, $\lambda^{(1/2)}$ satisfy Eq.~\eqref{eq:igg_two_cocycle} up to a phase:
\begin{align}
    \lambda^{(a)}(g_1,g_2)\lambda^{(a)}(g_1g_2,g_3)=\omega(g_1,g_2,g_3)^{(-1)^a}{}^{g_1}\lambda^{(a)}(g_2,g_3)\lambda^{(a)}(g_1,g_2g_3).
    \label{eq:def_omega}
\end{align}

We now apply to above formalism to the bosonic correlated TI discussed in the main text, with symmetry group $\mathbb{Z}_2\times U(1)$.
Here, $\mathcal{H}^{3}(U(1)\times\ZZ_2,U(1))=\ZZ\times \ZZ_2^2$: the $\ZZ$ part arise from $\mathcal{H}^{3}(U(1),U(1))$, corresponds to bosonic integer quantum Hall.
The first $\ZZ_2$ factor arise from $\mathcal{H}^{3}(\ZZ_2,U(1))$, and the second $\ZZ_2$ from $\mathcal{H}^{2}(\ZZ_2,\ZZ)$ according to the K{\"u}nneth formula, corresponding to the correlated TI in the main text.
Denoting $W(h)$ and $W(\theta)=\ee^{\ii\theta n}$ as the $\mathbb{Z}_2$ and $U(1)$ symmetry gauge transformation, the tensor equations captures such cohomology are
\begin{equation}
\begin{aligned}
    [W(h)]^2&= \hat{1}\\
    W(h)\circ n_c &=n+m^{(1)}+m^{(2)}\\
    W(h)\circ m^{(a)}&=-m^{(a)}+ (-\hat{1})^{a}
\end{aligned}
\label{eq:boson_tensor}
\end{equation}
where $m^{(1/2)}$ is integer-valued diagonal matrices generating $U(1)$ local $\igg$.
$(-\hat{1})^a$ in the last line signals the nontrivial cocycle.
We will show that this additional $(-\hat{1})^a$ in tensor equations leads to boundary anomaly in Section~\ref{subapp:bdry_anomaly}. 

\subsection{Tensor equations for fermionic topological insulators}
The classification and representation of fermionic topological insulators using fermionic PEPS is similar to the bosonic case, where details are presented in \cite{xu2024unveiling}.
The main difference is due to the possible minus signs from swapping $W(g)$'s, as $W(g)$ may not have fixed fermi parity.
For a time-reversal invariant fermionic correlated TI, the symmetry group is generated by the fermion number operator $n_f$, and time-reversal $\mathcal{T}$, where
\begin{align}
    \mathcal{T}^2=F\equiv\ee^{\ii\pi n_f}~, \quad
    \mathcal{T}\cdot n_f\cdot\mathcal{T}^{-1}=n_f\,.
\end{align}
Tensor equations for this phase are\cite{ma2024variational} 
\begin{equation}
\begin{aligned}
    W(\TT)\circ n_{f}&=n_f+m^{(1)}+m^{(2)}\\
    W(\TT)\KK\circ m^{(a)}&=-m^{(a)}+(-1)^{a}\\
    W(F)&=\ee^{\ii\frac{\pi}{2}(m^{(1)}+m^{(2)})^2}\cdot W(\TT) W^{*}(\TT)
    \label{eq:tensor_eqs_fermion}
\end{aligned}
\end{equation}
where $m^{(a)}$ generates local $U(1)$ $\igg$.
The local tensors satisfy the following symmetry condition
\begin{equation}
\begin{aligned}
\left(\bigotimes_{f,\alpha}W_{(j\alpha)}(\mathcal{T})\right) \cdot T_j=U_{j}(\mathcal{T})\cdot T^{*}_j~, \quad
\left( n_{f,j}+\sum_{\alpha} n_{f,(j\alpha)} \right)\cdot T_j=0\\
B^*_{jj'}\cdot \left( W^{\dagger}_{(j\alpha)}(\mathcal{T})\otimes W^{\dagger}_{(j'\beta)}(\mathcal{T}) \right)=B_{jj'}~,\quad 
B_{jj'}\cdot \left( n_{f,(j\alpha)}+n_{f,(j'\beta)} \right)=0
\end{aligned}
\end{equation}
as well as the constraints imposed by local IGG as in Fig.~\ref{fig:tensor_eq}(c). 

\subsection{Anomalous boundary theory from tensor equations}\label{subapp:bdry_anomaly}
The anomalous boundary symmetry of TI is encoded in the fusion rule of the symmetry defects\cite{levin2012}.
For the bosonic TI protected by $U(1)\times\ZZ_2$, fusion of two $\ZZ_2$ symmetry defects would give a positive/negative $U(1)$ charge.
We now derive this anomaly property from the tensor equations of PEPS.

For a PEPS on a region $A$ with an open boundary condition, denoted as $T_A$, the effective boundary Hilbert space are obtained from the boundary virtual legs $\HH_{\partial A}=\otimes_{j\in\partial A} \HH_j$.
More specifically, when $T_A$ is viewed as as a linear map from the boundary virtual legs $\HH_{\partial A}$ to the bulk physical legs $\HH_{A}$, the physical boundary Hilbert space is the image of $\TT_A$, and is isomorphic to $\HH_{\partial A}/\ker{T_A}$.
For the correlated TI, this space is defined by the projector $P_e=\prod_{j\in\partial A}P_{j+\frac{1}{2}}$, where $P_{j+\frac{1}{2}}[m^{(2)}_{j}+m^{(1)}_{j+1}]=0$.

Two $h$-symmetry defects at boundary site $1$ and $l+1$ are created by the following string operator acting on $\HH_{\partial A}$
\begin{align}
P_{e}\cdot w_1\left(\bigotimes^{l}_{j=2} W_j(h) \right) w_{l+1} \cdot P_e
\end{align}
where $w_1$ and $w_{l+1}$ are local excitations at ends of the string.
To simplify the derivation, $w_1$ and $w_{l+1}$ are chosen to satisfy the following condition:
\begin{align}
    \left[ P_e, w_1 \left( \bigotimes^{l}_{j=2} W_j(h) \right) w_{l+1} \right]=0
\end{align}
Combine with Eq.~\eqref{eq:boson_tensor}, we have
\begin{equation}
\begin{aligned}
    w_1\circ m^{(2)}_{1}&=-m^{(2)}_{1}+\hat{1}~,
    & w_1\circ m^{(1)}_{1}&=m^{(1)}_{1}\\
     w_{l+1}\circ m^{(2)}_{l+1}&=m^{(2)}_{l+1}~,
     & w_{l+1}\circ m^{(1)}_{l+1}&=-m^{(1)}_{l+1}-\hat{1}
\end{aligned}
    \label{eq:flux_flip}
\end{equation}
By further imposing the $U(1)$ conservation rule on the $\mathbb{Z}_2$ symmetry defects, we obtain 
\begin{equation}
    w_1\circ n_{1}=n_1+m^{(2)}_{1}~,\quad 
    w_{l+1}\circ n_{l+1}=n_{l+1}+m^{(1)}_{l+1}
    \label{eq:flux_charge}
\end{equation}
Combine Eq.~\eqref{eq:boson_tensor}, Eq.~\eqref{eq:flux_flip} and Eq.~\eqref{eq:flux_charge}, we have 
\begin{align}
   w^2_1\circ n= n+\hat{1} \quad w^2_{l+1}\circ n=n-\hat{1}
\end{align}
Namely, fusion of two $\mathbb{Z}_2$ symmetry defects gives a $U(1)$ charge.

Following a similar method, one can also show that in the QSH phase, the fusion of two $\TT$ fluxes gives a fermion mode, where the details are presented in Ref.~\cite{ma2024variational}.

\section{Disorder parameter in bosonic correlated TI}
\label{app:bcTI}
\subsection{RDM as a 1D intrinsic average SPT phase}
As stated in the main text, the RDM on the entanglement space $\rho_{\partial A}$ holds the following weak symmetry:
\begin{equation}   
 \rho_{\partial A}= W_{\partial A}(h)\cdot \rho_{\partial A}\cdot W^{\dagger}_{\partial A}(h)
 = W_{\partial A}(\theta)\cdot \rho_{\partial A}\cdot W^{\dagger}_{\partial A}(\theta)
\end{equation}
Further, due to the local constraint imposed by plaquette $\igg$, $\rho_{\partial A}$ in fact holds a strong $U(1)$ symmetry, where
\begin{equation}
 \rho_{\partial A}=D_{\partial A}(\theta)\cdot \rho_{\partial A}
 =\rho_{\partial A}\cdot D_{\partial A}(\theta)~,\quad
 \text{where }~D_{\partial A}(\theta)\equiv\prod_j \ee^{\ii \theta (m^l_j-m^r_j)}
\end{equation}
The algebra relations between $W_{\partial A}(h)$, $W_{\partial A}(\theta)$, and $D_{\partial A}(\theta)$ are derived from Eq.~\eqref{eq:boson_tensor} by identify $m_D=m^l-m^r$ and $m^{(1/2)}_{\lambda}=\pm m^{l/r}$. 
Such relations lead to a non-vanishing string order parameter: 
\begin{align}
    \lim_{\abs{L}\to \infty} \Tr \left[  \rho_{\partial A}\cdot \ee^{\ii\pi m_1^r} \left( \prod^{L-1}_{j=2} D_j(\pi) \right) \ee^{\ii \pi m^l_{L}}   \right]=1
\end{align}
Here, $\prod^{L-1}_{j=2} D_j(\pi)$ is a $\pi$-flux line of strong $U(1)$ symmetry and $\ee^{\pm\ii\pi m^{l/r}}$ carries the weak $\ZZ_2$ symmetry charge:
\begin{align}
    W(h) \cdot\ee^{\ii\pi m^{a}_j} = (-)\,\ee^{\pm\ii\pi m^{a}_j}\cdot W(h)~,\quad a=l,r
\end{align}
The non-vanishing string order parameter indicates that $\rho_{\partial A}$ belongs to an intrinsic average SPT protected by the weak $\ZZ_2$ symmetry and the strong $U(1)$ symmetry. 

\subsection{Disorder parameter from MPDO}
As shown in the main text, the RDM in the entanglement space of the correlated TI is well approximated as an injective MPDO
\begin{equation}
    \rho_{\partial A}
    =~
\begin{tikzpicture}[scale=1,baseline={([yshift=-2pt]current bounding box.center)}]
\def\Mbox#1#2{
    \draw (#1-\boxespacing,#2-\boxespacing) -- ++(0,2*\boxespacing) -- ++(2*\boxespacing,0) -- ++(0,-2*\boxespacing)-- cycle;
    \node[anchor=center] at (#1,#2) {$M$}; 
    \draw (#1,#2+\boxespacing)-- ++(0,0.25);
    \draw (#1,#2-\boxespacing)-- ++(0,-0.25);
}
\Mbox{0}{0}

\Mbox{0.8}{0}
\Mbox{2.25}{0}
\draw (0-\boxespacing,0)-- ++(-0.25,0);
\draw[dotted] (0-\boxespacing,-0.1) -- ++ (-0.25,0);
\draw (0-\boxespacing-0.25,0) arc (90:270:0.05);
\draw (0+\boxespacing,0)-- (0.8-\boxespacing,0);
\draw[dotted] (0.8+\boxespacing,-0.1)--++(0.2,0);
\draw (0.8+\boxespacing,0)--++(0.2,0);
\node[black, anchor=center] at (1.55,0) {$\dots$};
\draw(2.25-\boxespacing,0)--++(-0.1,0);
\draw[dotted] (2.25-\boxespacing,-0.1)--++(-0.1,0);
\draw (2.25+\boxespacing,0)-- ++ (0.25,0);
\draw[dotted] (0.8-\boxespacing,-0.1)--(0+\boxespacing,-0.1);
\draw[dotted] (2.25+\boxespacing+0.25,-0.1)--++(-0.25,0);
\draw (2.25+\boxespacing+0.25,0) arc (90:-90:0.05);
\end{tikzpicture}
    \label{eq:rho_mpdo_sm}\,.
\end{equation}
where the injectivity signals short-range correlations.

For large subsystem where $\abs{\partial A}\gg 1$, the disorder parameter admits approximation
\begin{align}
\lrangle{U_A(\theta)}=\Tr([T(\theta)]^{\abs{\partial A}}) \approx  \sum_{l=1}^{k}[\lambda_{l}(\theta)]^{\abs{\partial A}}\,.
\label{eq:disorder_param_transfer_mat_eigenval}
\end{align}
where $T(\theta)\equiv\sum_{\gamma\delta}\left( \sum_{\alpha\beta}[W(\theta)]_{\beta\alpha} M_{\alpha\beta;\gamma\delta} \right) \vket{\gamma}\vbra{\delta}\equiv\vcenter{\hbox{\includegraphics[height=3.5em]{fig/trans_matrix.pdf}}}$ is the $\theta$-twisted transfer matrix, and $\{\lambda_l(\theta)|l=1,\ldots,k\}$ are its $k$-degenerate dominant eigenvalues with the same modulus.
For the untwisted $T\equiv T(\theta=0)$, $k=1$ due to the injectivity condition.
Behavior of disorder parameter now reduces to the spectral flow of $T(\theta)$, which can be further constrained by symmetries of $T(\theta)$.

To get symmetries of $T(\theta)$, we investigate the symmetry properties of local tensors, as shown in Eq.~(13) and Eq.~(14) in the main text (list below):
\begin{equation}
    \begin{tikzpicture}[scale=1,baseline={([yshift=-2pt]current bounding box.center)}]
\def\Mbox#1#2{
    \draw (#1-\boxespacing,#2-\boxespacing) -- ++(0,2*\boxespacing) -- ++(2*\boxespacing,0) -- ++(0,-2*\boxespacing)-- cycle;
    \node[anchor=center] at (#1,#2) {$M$}; 
    \draw[thin] (#1,#2+\boxespacing)-- ++(0,0.25);
    \draw[thin] (#1,#2-\boxespacing)-- ++(0,-0.25);
    \draw[thin] (#1-\boxespacing,#2)-- ++(-0.25,0);
    \draw[thin] (#1+\boxespacing,#2)-- ++(0.25,0);
}

\Mbox{0.3}{0}
\node[anchor=west] at (0+\boxespacing+0.6,-0.01) {$=$};

\Mbox{2}{0}
\node[red,scale=1.5,anchor=center] at (2,0+\boxespacing+0.15) {$\cdot$};
\node[red, scale=1.5,anchor=center] at (2+\boxespacing+0.15,0) {$\cdot$};
\node[red, scale=1.5,anchor=center] at (2-\boxespacing-0.15,0) {$\cdot$};
\node[red,scale=0.6,anchor=east] at (2,0+\boxespacing+0.15) {$W(h)$};
\node[red, scale=0.6,anchor=west] at (2+\boxespacing,0+0.15) {$V^{\dagger}(h)$};
\node[red, scale=0.6,anchor=east] at (2-\boxespacing,0-0.15) {$V(h)$};
\node[red,scale=1.5,anchor=center] at (2,0-\boxespacing-0.15) {$\cdot$};
\node[red,scale=0.6,anchor=west] at (2+0.05,0-\boxespacing-0.15) {$W^{\dagger}(h)$};

\begin{scope}[xshift=2cm]
\node[anchor=west] at (0+\boxespacing+0.6,-0.01) {$=$};

\Mbox{2}{0}
\node[red,scale=1.5,anchor=center] at (2,0+\boxespacing+0.15) {$\cdot$};
\node[red, scale=1.5,anchor=center] at (2+\boxespacing+0.15,0) {$\cdot$};
\node[red, scale=1.5,anchor=center] at (2-\boxespacing-0.15,0) {$\cdot$};
\node[red,scale=0.6,anchor=east] at (2,0+\boxespacing+0.15) {$W(\theta)$};
\node[red, scale=0.6,anchor=west] at (2+\boxespacing,0+0.15) {$V^{\dagger}(\theta)$};
\node[red, scale=0.6,anchor=east] at (2-\boxespacing,0-0.15) {$V(\theta)$};
\node[red,scale=1.5,anchor=center] at (2,0-\boxespacing-0.15) {$\cdot$};
\node[red,scale=0.6,anchor=west] at (2+0.05,0-\boxespacing-0.15) {$W^{\dagger}(\theta)$};
\end{scope}
\end{tikzpicture}
\label{eq:tensor_weak_sym_sm}
\end{equation}
as well as
\begin{equation}
    \begin{tikzpicture}[scale=1,baseline={([yshift=-5pt]current bounding box.center)}]
\def\Mbox#1#2{
    \draw (#1-\boxespacing,#2-\boxespacing) -- ++(0,2*\boxespacing) -- ++(2*\boxespacing,0) -- ++(0,-2*\boxespacing)-- cycle;
    \node[anchor=center] at (#1,#2) {$M$}; 
    \draw[thin] (#1,#2+\boxespacing)-- ++(0,0.25);
    \draw[thin] (#1,#2-\boxespacing)-- ++(0,-0.25);
    \draw[thin] (#1-\boxespacing,#2)-- ++(-0.25,0);
    \draw[thin] (#1+\boxespacing,#2)-- ++(0.25,0);
}

\Mbox{0.3}{0}
\node[anchor=west] at (0+\boxespacing+0.6,-0.01) {$=$};
\node[red,scale=1.5,anchor=center] at (0.3,0+\boxespacing+0.15) {$\cdot$};
\node[red,scale=0.6,anchor=west] at (0.3,0+\boxespacing+0.15) {$m^r$};

\Mbox{2}{0}
\node[red, scale=1.5,anchor=center] at (2+\boxespacing+0.15,0) {$\cdot$};
\node[red, scale=0.6,anchor=west] at (2+\boxespacing,0+0.15) {$m^u$};

\begin{scope}[xshift=1.9cm]
\node[anchor=west] at (0+\boxespacing+0.6,-0.01) {$,$};

\Mbox{2}{0}
\node[red,scale=1.5,anchor=center] at (2,0+\boxespacing+0.15) {$\cdot$};
\node[red,scale=0.6,anchor=east] at (2,0+\boxespacing+0.15) {$m^l$};
\end{scope}

\begin{scope}[xshift=3.6cm]
\node[anchor=west] at (0+\boxespacing+0.6,-0.01) {$=$};

\Mbox{2}{0}
\node[red, scale=1.5,anchor=center] at (2-\boxespacing-0.15,0) {$\cdot$};
\node[red, scale=0.6,anchor=east] at (2-\boxespacing,0+0.15) {$m^u$};
\end{scope}

\begin{scope}[yshift=-1.5cm]
\Mbox{0.3}{0}
\node[anchor=west] at (0+\boxespacing+0.6,-0.01) {$=$};
\node[red,scale=1.5,anchor=center] at (0.3,0-\boxespacing-0.15) {$\cdot$};
\node[red,scale=0.6,anchor=west] at (0.3,0-\boxespacing-0.15) {$m^r$};

\Mbox{2}{0}
\node[red, scale=1.5,anchor=center] at (2+\boxespacing+0.15,0) {$\cdot$};
\node[red, scale=0.6,anchor=west] at (2+\boxespacing,0-0.15) {$m^d$};

\begin{scope}[xshift=1.9cm]
\node[anchor=west] at (0+\boxespacing+0.6,-0.01) {$,$};

\Mbox{2}{0}
\node[red,scale=1.5,anchor=center] at (2,0-\boxespacing-0.15) {$\cdot$};
\node[red,scale=0.6,anchor=east] at (2,0-\boxespacing-0.15) {$m^l$};
\end{scope}

\begin{scope}[xshift=3.6cm]
\node[anchor=west] at (0+\boxespacing+0.6,-0.01) {$=$};

\Mbox{2}{0}
\node[red, scale=1.5,anchor=center] at (2-\boxespacing-0.15,0) {$\cdot$};
\node[red, scale=0.6,anchor=east] at (2-\boxespacing,0-0.15) {$m^d$};
\end{scope}
\end{scope}
\end{tikzpicture}
\label{eq:tensor_strong_plq_sym_sm}
\end{equation}
The algebraic relations between $W(h)$, $W(\theta)$ and $m^{r/l}$ follows Eq.~\eqref{eq:boson_tensor}:
\begin{align}
    W(\theta)=\ee^{\ii \theta n_c},\quad W(h)\circ n_c=n_c+m^{r}-m^{l}, \quad W(h)\circ m^{r/l}=\hat{1}-m^{r/l}
\end{align}
leading to the same algebra relation for symmetries on the internal legs of MPDO:
\begin{equation}
    V(\theta)=\ee^{\ii\theta n}, \quad V(h)\circ n=n+m^u-m^d,\quad V^{a}(h)\circ m^{u/d}=\hat{1}-m^{u/d}.
    \label{eq:V_m_relation_sm}
\end{equation}

Symmetries of $T(\theta)$ are derived from from those of the $M$.
Note that from Eq.~\eqref{eq:tensor_strong_plq_sym_sm}, $(m^u-m^d)\cdot T(\theta)=T(\theta)\cdot (m^u-m^d)=0$, therefore, we could focus on $m^{u}$, and igonore its superscript for simplicity. 
From Eq.~\eqref{eq:tensor_weak_sym_sm} and Eq.~\eqref{eq:V_m_relation_sm}, we have 
\begin{equation}
    \begin{tikzpicture}
\def\Mbox#1#2{
    \draw (#1-\boxespacing,#2-\boxespacing) -- ++(0,2*\boxespacing) -- ++(2*\boxespacing,0) -- ++(0,-2*\boxespacing)-- cycle;
    \node[anchor=center] at (#1,#2) {$M$}; 
    \draw[thin] (#1,#2+\boxespacing)-- ++(0,0.35);
    \draw[thin] (#1,#2-\boxespacing)-- ++(0,-0.25);
    \draw[thin] (#1-\boxespacing,#2)-- ++(-0.25,0);
    \draw[thin] (#1+\boxespacing,#2)-- ++(0.25,0);
    \draw [my dash,thin] (#1+0.1,#2+\boxespacing)-- ++(0,0.35);
    \draw [my dash,thin] (#1+0.1,#2-\boxespacing)-- ++(0,-0.25);
    \draw[thin] (#1,#2+\boxespacing+0.35) arc (-180:-360:0.05);
    \draw[thin] (#1,#2-\boxespacing-0.25) arc (-180:0:0.05);
}

\Mbox{0}{0}
\draw[thin] (0-\boxespacing-0.15,0)--++(-0.2,0);
\draw[thin] (0+\boxespacing+0.15,0)--++(0.2,0);
\node [red, scale=1.5, anchor=center] at (0,+\boxespacing+0.15) {$\cdot$};
\node [red,anchor=east,scale=0.6] at (0,+\boxespacing+0.15) {$W(\theta)$};
\node[anchor=west] at (0+\boxespacing+0.5,-0.01) {$=$};
\node[red,anchor=center,scale=1.5] at (-\boxespacing-0.25,0) {$\cdot$};
\node[red,anchor=north,scale=0.6] at (-\boxespacing-0.28,0) {$V(h)$};
\node[red,anchor=center,scale=1.5] at (+\boxespacing+0.25,0) {$\cdot$};
\node[red,anchor=north,scale=0.6] at (+\boxespacing+0.28,0) {$V^{\dagger}(h)$};
\Mbox{2}{0}
\node [red, scale=1.5, anchor=center] at (2,+\boxespacing+0.25) {$\cdot$};
\node [red,anchor=east,scale=0.6] at (2,+\boxespacing+0.25) {$W(\theta)$};
\node [red, scale=1.5, anchor=center] at (2,+\boxespacing+0.1) {$\cdot$};
\node [red,anchor=west,scale=0.6] at (2,+\boxespacing+0.1) {$W^{\dagger}(h)$};
\node [red, scale=1.5, anchor=center] at (2,-\boxespacing-0.15) {$\cdot$};
\node [red,anchor=east,scale=0.6] at (2,-\boxespacing-0.15) {$W(h)$};
\node[anchor=west] at (2+\boxespacing+0.5,0) {$=$};
\Mbox{4}{0}
\node [red, scale=1.5, anchor=center] at (4,+\boxespacing+0.25) {$\cdot$};
\node [red,anchor=east,scale=0.6] at (4,+\boxespacing+0.25) {$W(\theta)$};
\node [red, scale=1.5, anchor=center] at (4,+\boxespacing+0.1) {$\cdot$};
\node [red,anchor=west,scale=0.6] at (4,+\boxespacing+0.1) {$\exp(\ii\theta(m^l-m^r))$};
\node[anchor=west] at (4+\boxespacing+0.7,0) {$=$};
\Mbox{7}{0}
\node [red, scale=1.5, anchor=center] at (7,+\boxespacing+0.25) {$\cdot$};
\node [red,anchor=east,scale=0.6] at (7,+\boxespacing+0.25) {$W(\theta)$};
\draw[thin] (7-\boxespacing-0.25,0)-- ++(-0.15,0);
\draw[thin] (7+\boxespacing+0.25,0)-- ++(0.15,0);
\node[red,scale=1.5,anchor=center] at (7-\boxespacing-0.2,0) {$\cdot$};
\node[red,scale=0.6,anchor=north east] at (7-\boxespacing-0.2,0) {$\ee^{-\ii\theta m}$};
\node[red,scale=1.5,anchor=center] at (7+\boxespacing+0.3,0) {$\cdot$};
\node[red,scale=0.6,anchor=north west] at (7+\boxespacing+0.3,0) {$\ee^{\ii\theta m}$};
\node[anchor=west] at (7+\boxespacing+0.7,0) {$=$};
\Mbox{10}{0}
\node [red, scale=1.5, anchor=center] at (10,+\boxespacing+0.25) {$\cdot$};
\node [red,anchor=east,scale=0.6] at (10,+\boxespacing+0.25) {$W(\theta)$};
\draw[thin] (10-\boxespacing-0.25,0)-- ++(-0.15,0);
\draw[thin] (10+\boxespacing+0.25,0)-- ++(0.15,0);
\node[red,scale=1.5,anchor=center] at (10-\boxespacing-0.2,0) {$\cdot$};
\node[red,scale=0.6,anchor=north east] at (10-\boxespacing-0.2,0) {$\ee^{-\ii\theta m}$};
\node[red,scale=1.5,anchor=center] at (10+\boxespacing+0.3,0) {$\cdot$};
\node[red,scale=0.6,anchor=north west] at (10+\boxespacing+0.3,0) {$\ee^{\ii\theta m}$};
\end{tikzpicture}
\end{equation}
Therefore, $\ee^{-\ii \theta m}V(h)$ is a symmetry of $T(\theta)$.
From Eq.~\eqref{eq:V_m_relation_sm}, we conclude that $\left( \ee^{-\ii \theta m}V(h) \right)^2=\ee^{-\ii\theta}$.
By absorbing the phase factor, we define the $\ZZ_2$ symmetry action generated by 
\begin{equation}
        \mathcal{V}_{\theta}\equiv \exp\left[-\ii\theta(m-\frac{1}{2}) \right]V(h)~,\quad \text{with }\mathcal{V}_{\theta+2\pi}=- \mathcal{V}_{\theta}
        \label{eq:twisted_transfer_mat_zn_sym}
\end{equation}

We now investigate the spectral flow of $T(\theta)$.
Let $v^r_0$ be the unique (right) dominant eigenstate of $T$, with eigenvalue $\lambda_0=1$ and quantum number $s_0\in\{1,-1\}$ under $\mathcal{V}^u_{\theta=0}$.
As $\theta$ flows 0 to $2\pi$,  continuity implies the quantum number remain fixed, $s_0(\theta)=s_0$.
However, since $\mathcal{V}_0^a=-\mathcal{V}_{2\pi}^a$, the evolved eigenvector $v_0^r(2\pi)$ must have quantum number $-s_0$ under $\mathcal{V}_0^a$,
Therefore, $v^r_0(2\pi)$ is a distinct eigenvector of $T$ from $v^r_0$ with eigenvalue $\lambda_1$ whose modulus. 
This enforces a crossing the magnitude of their eigenvalues, $\abs{\lambda_0(\theta)}$ and $\abs{\lambda_1(\theta)}$, as $\theta$ varies, and hence leads to non-smooth behavior of $\abs{\lrangle{U_A(\theta)}}$ according to Eq.~\eqref{eq:disorder_param_transfer_mat_eigenval}.

We now impose the Hermiticity condition for $\rho_{\partial A}$, which gives additional constraint on local tensors as
\begin{equation}
    \begin{tikzpicture}[scale=1,baseline={([yshift=-3pt]current bounding box.center)}]
\def\Mbox#1#2{
    \draw (#1-\boxespacing,#2-\boxespacing) -- ++(0,2*\boxespacing) -- ++(2*\boxespacing,0) -- ++(0,-2*\boxespacing)-- cycle;
    
    \draw[thin] (#1,#2+\boxespacing)-- ++(0,0.25);
    \draw[thin] (#1,#2-\boxespacing)-- ++(0,-0.25);
    \draw[thin] (#1-\boxespacing,#2)-- ++(-0.3,0);
    \draw[thin] (#1+\boxespacing,#2)-- ++(0.3,0);
}
\node[black, scale=0.7,anchor=south] at (0.3,0+\boxespacing+0.25) {$\alpha$};
\node[black, scale=0.7, anchor=north] at (0.3,0-\boxespacing-0.2) {$\beta$};
\Mbox{0.3}{0}
\node[anchor=center] at (0.3,0) {$M$}; 
\node[anchor=west] at (0+\boxespacing+0.6,-0.01) {$=$};
\node[black, scale=0.7,anchor=south] at (2,0+\boxespacing+0.2) {$\beta$};
\node[black, scale=0.7, anchor=north] at (2,0-\boxespacing-0.2) {$\alpha$};
\Mbox{2}{0}
\node[anchor=center] at (2,0) {$M^{*}$}; 
\node[red, scale=1.5,anchor=center] at (2+\boxespacing+0.15,0) {$\cdot$};
\node[red, scale=1.5,anchor=center] at (2-\boxespacing-0.15,0) {$\cdot$};
\node[red, scale=0.6,anchor=west] at (2+\boxespacing,0-0.15) {$Q^{\dagger}$};
\node[red, scale=0.6,anchor=east] at (2-\boxespacing,0-0.15) {$Q$};
\end{tikzpicture}
\label{eq:tensor_hermiticity}
\end{equation}
for some gauge transformation $Q$, satisfying
\begin{equation}
    (Q\KK)^2=1~,\quad
    [Q\KK,V(h)]=[Q\KK,V(\theta)]=0~,\quad
    Q\KK \cdot m^u=m^d\cdot Q\KK~.
    \label{eq:hermition_group_relation}
\end{equation}
Therefore, when applying on $T(\theta)$, we have
\begin{equation}
    Q\KK\cdot T(\theta)=T(-\theta)\cdot Q\KK
    \label{eq:T_hermicity}
\end{equation}
In particular, $T(n\pi)$ is symmetric under the anti-unitary operation $Q\KK$.
Let $v^r(\theta)$ be a right eigenstate of $T(\theta)$, and with quantum number $s$ under $\VV_\theta$.
At $\theta=\pi$, we have
\begin{equation}
    V_\pi\cdot Q\KK \cdot v_\pi^r
    =Q\KK\cdot V_{-\pi} \cdot v_\pi^r
    =-Q\KK\cdot V_{\pi} \cdot v_\pi^r
    =(-s) Q\KK \cdot  v_\pi^r
\end{equation}
where we use $Q\KK\cdot \VV_\theta= \VV_{-\theta}\cdot Q\KK$ (derived from Eq.~\eqref{eq:hermition_group_relation}) in the first equation.
Hence, $v_\pi^r$ and $Q\KK\cdot v_\pi^r$ are distinct eigenstates of $T(\pi)$, whose eigenvalues are conjugate to each other.

\begin{figure}[h]
    \centering
    \includegraphics[width=0.8\linewidth]{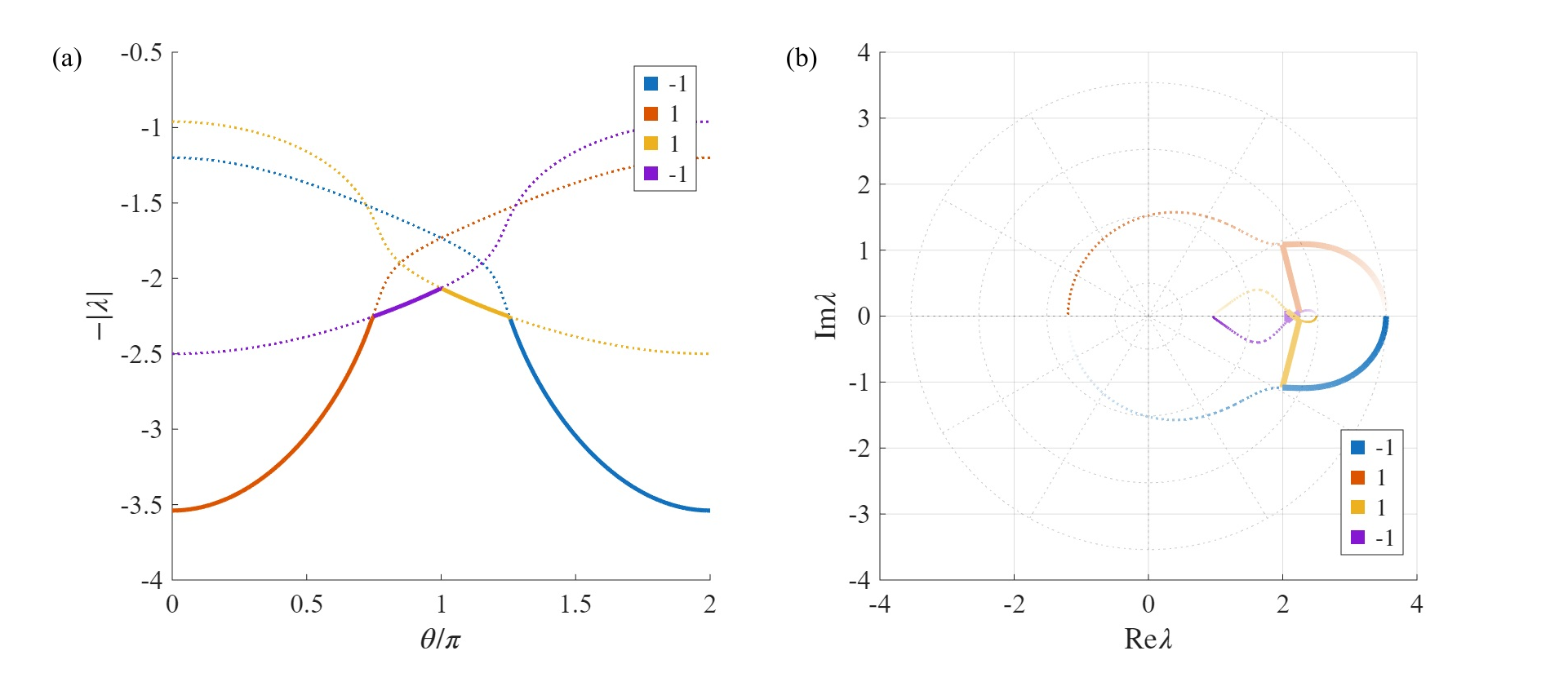}
    \caption{The modulus (a) and the complex plane trajectories of eigenvalues of $T(\theta)$ with $A=\bigl(\begin{smallmatrix} 1.4 & 1.2 \\ 1 & 1.5 \end{smallmatrix}\bigr)$ and $B=\bigl(\begin{smallmatrix} -2.6 & 1.9 \\ -1 & 1 \end{smallmatrix}\bigr)$. 
    The solid lines represent the eigenvalues with largest modulus.
    Here, three singular points are present.}
    \label{fig:sf_2}
\end{figure}

Therefore, the simplest behavior of disorder parameter in this phase is presented in the main text, where $f(\theta)$ is smooth everywhere except at $\theta=\pi$.
However, other cases may also arise.
As a demonstration, let's consider the following $T(\theta)$:
\begin{align}
    T(\theta)=\begin{pmatrix}
       A & B e^{i\theta} \\
        B & A
    \end{pmatrix}
\end{align}
Where $A$ and $B$ are $n\times n$ real matrices.
Let $\VV_{\theta}=(X\otimes \hat{1}_{n\times n}) \ee^{\ii \theta(m-\frac{1}{2})}$, $m=\hat{0}_{n\times n}\oplus \hat{1}_{n\times n}$ and $Q=\hat{1}_{2n\times 2n}$, $T(\theta)$ then satisfy the desired constraints.
By choosing proper $A$ and $B$, we may obtain three singular points, as shown in Fig.~\ref{fig:sf_2}.

\subsection{Numerical details for bosonic correlated TI}
The bosonic PEPS wavefunction used in the main text is defined in a honeycomb lattice. 
The building block includes a bond tensor ${B}$ and site tensor ${T}_{u/v}=\sum c_l {t}^{l}_{u/v}$, where ${t}_l$ is a set of tensor basis that satisfies Eq.~\eqref{eq:boson_tensor}.
Graphically, these local tensors are represented as
\begin{align}
    \adjincludegraphics[valign=c,scale=1]{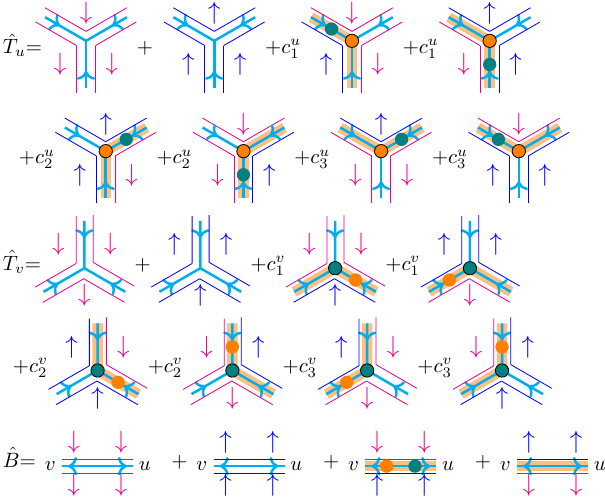}
\end{align}
Here, the teal/orange vertices denote negative/positive charged bosons, and the arrows represent Ising spins.
Here, Ising spins live on both in the physical legs and internal legs.
The local tensor is invariant under $W(h)=\vket{\uparrow\uparrow}\vbra{ \downarrow\downarrow}+\vket{\uparrow\downarrow} \vbra{\downarrow\uparrow}+\vket{\downarrow\uparrow}\vbra{\uparrow\downarrow}+\vket{\downarrow\downarrow}\vbra{ \uparrow\uparrow}$.
Here, the arrows are ordered clockwise for the $u/v$ sublattice tensor.
The charge operator is set to be $n_{c,u}=-\vket{\uparrow\downarrow}\vbra{\uparrow\downarrow}$, $n_{c,v}=\vket{\uparrow\downarrow} \vbra{\uparrow\downarrow}$.
The generator for local plaquette $\igg$ $m^{(1)}=-\vket{\uparrow\downarrow}\vbra{\uparrow\downarrow}-\vket{\uparrow\uparrow}\vbra{ \uparrow\uparrow}$ and $m^{(2)}_{\lambda}=\vket{\downarrow\uparrow} \vbra{\downarrow\uparrow}+\vket{\uparrow\uparrow}\vbra{\uparrow\uparrow}$. 
The fixed-point wavefunction is obtained by setting all $c^{u/v}_{l}=1$.

\section{Fermionic correlated TI with time-reversal symmetry}
\label{app:fcTI}
\subsection{Disorder parameter from MPDO}
The reduced density matrix of fermionic correlated TI with time-reversal symmetry can also be approximated by the injective MPO $\rho_{\partial A}$. Since $\rho_{\partial A}$ is parity even, we can assume the local tensor $M$ to be parity even without loss of generality.

$\rho_{\partial A}$ possesses weak and strong symmetries
\begin{align}
    \rho_{\partial A}=W_{\partial A}(\mathcal{T}) \rho^{*}_{A} W^{\dagger}_{\partial
     A}(\TT)\,,\quad \rho_{\partial A}=\Lambda^{(2)}(\theta)\Lambda^{(1)}(\theta)
\end{align}
Here, $\Lambda^{(a)}(\theta)=\ee^{\ii \theta m^{(a)}}$. They can be lifted to symmetries of the local tensor $M$ as in Eq.~\eqref{eq:tensor_weak_sym_sm} and \eqref{eq:tensor_strong_plq_sym_sm}. However, note that $W(\mathcal{T})\circ n_f\neq n_f$, from Ref.~\cite{xu2024unveiling}, it turns out that
\begin{align}
    W_{\partial A}(\mathcal{T})=\bigotimes_{j\in\partial A}{}_f  
  \ee^{\ii \pi\xi(j)}W_j(\mathcal{T})
\end{align}
where $\xi(j)$ is an integer-valued function. However, one can always attribute the $\ee^{\ii \pi \xi(j)}$ part to $W_{\partial A}^{-1}$ acting on ${T}_{\bar{A}}$. Thus we set
\begin{equation}
    W_{\partial A}(\mathcal{T})=\bigotimes^{|\partial A|}_{j=1}{}_f W_{j}(\mathcal{T})\,
\end{equation}
and 
\begin{equation}
\begin{tikzpicture}
\def\Mbox#1#2{
    \draw (#1-\boxespacing,#2-\boxespacing) -- ++(0,2*\boxespacing) -- ++(2*\boxespacing,0) -- ++(0,-2*\boxespacing)-- cycle;
    \node[anchor=center] at (#1,#2) {$M^{*}$}; 
    \draw (#1,#2+\boxespacing)-- ++(0,0.25);
    \draw (#1,#2-\boxespacing)-- ++(0,-0.25);
}
\def\Wg#1#2{
\node[anchor=center, red, scale=1.5] at (#1,#2) {$\cdot$};
\node[anchor=south, red,scale=0.6] at (#1,#2) {$W_{j}(\mathcal{T})$};
}
\def\Wdg#1#2{
\node[red,scale=1.5, anchor=center] at (#1,#2) {$\cdot$};
\node[red,anchor=north,scale=0.6] at (#1,#2) {$W^{\dagger}_j(\mathcal{T})$};
}
\node[anchor=east] at (-0.55,-0.1) {$\rho_{\partial A}= \otimes_{j} W_j(\mathcal{T})\cdot\rho^{*}_{\partial A}\cdot \otimes_f W^{\dagger}_{j}(\TT)=$};
\Mbox{0}{0}
\Wg{0}{0+\boxespacing+0.15}
\Wdg{0}{0-\boxespacing-0.15}
\Mbox{0.8}{0}
\Wg{0.8}{0+\boxespacing+0.15}
\Wdg{0.8}{0-\boxespacing-0.15}
\Mbox{2.25}{0}
\Wg{2.25}{0+\boxespacing+0.15}
\Wdg{2.25}{0-\boxespacing-0.15}
\draw (0-\boxespacing,0)-- ++(-0.25,0);
\draw[dotted] (0-\boxespacing,-0.1) -- ++ (-0.25,0);
\draw (0-\boxespacing-0.25,0) arc (90:270:0.05);
\draw (0+\boxespacing,0)-- (0.8-\boxespacing,0);
\draw[dotted] (0.8+\boxespacing,-0.1)--++(0.2,0);
\draw (0.8+\boxespacing,0)--++(0.2,0);
\node[black, anchor=center] at (1.55,0) {$\dots$};
\draw(2.25-\boxespacing,0)--++(-0.1,0);
\draw[dotted] (2.25-\boxespacing,-0.1)--++(-0.1,0);
\draw (2.25+\boxespacing,0)-- ++ (0.25,0);
\draw[dotted] (0.8-\boxespacing,-0.1)--(0+\boxespacing,-0.1);
\draw[dotted] (2.25+\boxespacing+0.25,-0.1)--++(-0.25,0);
\draw (2.25+\boxespacing+0.25,0) arc (90:-90:0.05);
\end{tikzpicture} 
\end{equation}

Thus, the local tensor $M$ thus has the following symmetries
\begin{equation}
    \begin{tikzpicture}
\def\Mbox#1#2#3{
    \draw (#1-\boxespacing,#2-\boxespacing) -- ++(0,2*\boxespacing) -- ++(2*\boxespacing,0) -- ++(0,-2*\boxespacing)-- cycle;
    \node[anchor=center] at (#1,#2) {$#3$}; 
    \draw[thin] (#1,#2+\boxespacing)-- ++(0,0.35);
    \draw[thin] (#1,#2-\boxespacing)-- ++(0,-0.25);
    \draw[thin] (#1-\boxespacing,#2)-- ++(-0.25,0);
    \draw[thin] (#1+\boxespacing,#2)-- ++(0.25,0);
}

\Mbox{0.3}{0}{M}
\node[red,scale=1.5,anchor=center] at (0.3-\boxespacing-0.15,0) {$\cdot$};
\node[red, scale=0.5,anchor=north] at (0.3-\boxespacing-0.2,0) {$V^{\dagger}(\TT)$};
\node[red,scale=1.5,anchor=center] at (0.3+\boxespacing+0.15,0) {$\cdot$};
\node[red, scale=0.5,anchor=north] at (0.3+\boxespacing+0.2,0) {$V(\TT)$};
\node[anchor=west] at (0+\boxespacing+0.6,-0.01) {$=$};

\Mbox{2}{0}{M^{*}}
\node[red,scale=1.5,anchor=center] at (2,0+\boxespacing+0.15) {$\cdot$};
\node[red,scale=0.5,anchor=east] at (2,0+\boxespacing+0.15) {$W(\TT)$};
\node[red,scale=1.5,anchor=center] at (2,0-\boxespacing-0.15) {$\cdot$};
\node[red,scale=0.5,anchor=west] at (2+0.05,0-\boxespacing-0.15) {$W^{\dagger}(\TT)$};

\Mbox{3.5}{0}{M}
\begin{scope}[xshift=3.1cm]
\node[scale=1,anchor=east] at (2-\boxespacing-0.3,0) {$=$};
\Mbox{2}{0}{M}
\node[red,scale=1.5,anchor=center] at (2,0+\boxespacing+0.15) {$\cdot$};
\node[red, scale=1.5,anchor=center] at (2+\boxespacing+0.15,0) {$\cdot$};
\node[red, scale=1.5,anchor=center] at (2-\boxespacing-0.15,0) {$\cdot$};
\node[red,scale=0.5,anchor=east] at (2,0+\boxespacing+0.15) {$W(\theta)$};
\node[red, scale=0.5,anchor=west] at (2+\boxespacing,0+0.15) {$V^{\dagger}(\theta)$};
\node[red, scale=0.5,anchor=east] at (2-\boxespacing,0-0.15) {$V(\theta)$};
\node[red,scale=1.5,anchor=center] at (2,0-\boxespacing-0.15) {$\cdot$};
\node[red,scale=0.5,anchor=west] at (2+0.05,0-\boxespacing-0.15) {$W^{\dagger}(\theta)$};
\node[anchor=west] at (2+\boxespacing+0.25,0) {$=$};
\Mbox{3.6}{0}{M}
\node[scale=1.5,red,anchor=center] at (3.6,\boxespacing+0.25) {$\cdot$};
\node[scale=0.6, red, anchor=west] at  (3.6,\boxespacing+0.25) {$\Lambda^{(1)}(\theta)$};
\node[scale=1.5,red,anchor=center] at (3.6-\boxespacing-0.15,0) {$\cdot$};
\node[scale=0.6, red, anchor=north] at  (3.6-\boxespacing-0.25,0) {$\iota(\theta)$};
\node[anchor=west] at (3.6+\boxespacing+0.25,0) {$=$};
\Mbox{5.1}{0}{M}
\node[scale=1.5,red,anchor=center] at (5.1,\boxespacing+0.25) {$\cdot$};
\node[scale=0.6, red, anchor=west] at  (5.1,\boxespacing+0.25) {$\Lambda^{(2)}(\theta)$};
\node[scale=1.5,red,anchor=center] at (5.1+\boxespacing+0.15,0) {$\cdot$};
\node[scale=0.6, red, anchor=north] at  (5.1+\boxespacing+0.25,0) {$\iota(-\theta)$};
\end{scope}
\end{tikzpicture}
\end{equation}
From the above equation and Eq.~\eqref{eq:tensor_eqs_fermion}, we have
\begin{equation}
    \begin{tikzpicture}
\def\Mbox#1#2#3{
    \draw (#1-\boxespacing,#2-\boxespacing) -- ++(0,2*\boxespacing) -- ++(2*\boxespacing,0) -- ++(0,-2*\boxespacing)-- cycle;
    \node[anchor=center] at (#1,#2) {$#3$}; 
    \draw[thin] (#1,#2+\boxespacing)-- ++(0,0.35);
    \draw[thin] (#1,#2-\boxespacing)-- ++(0,-0.25);
    \draw[thin] (#1-\boxespacing,#2)-- ++(-0.25,0);
    \draw[thin] (#1+\boxespacing,#2)-- ++(0.25,0);
}

\Mbox{-0.3}{0}{M}
\draw[thin] (-0.3-\boxespacing-0.15,0)--++(-0.2,0);
\draw[thin] (-0.3+\boxespacing+0.15,0)--++(0.2,0);
\node[red,scale=1.5,anchor=center] at (-0.3-\boxespacing-0.15,0) {$\cdot$};
\node[red, scale=0.5,anchor=north] at (-0.3-\boxespacing-0.55,0) {$V^{T}(\TT)V^{\dagger}(\TT)$};
\node[red,scale=1.5,anchor=center] at (-0.3+\boxespacing+0.25,0) {$\cdot$};
\node[red, scale=0.5,anchor=north] at (-0.3+\boxespacing+0.55,0) {$V(\TT) V^{*}(\TT)$};
\node[anchor=west] at (0+\boxespacing+0.6,-0.01) {$=$};

\Mbox{2}{0}{M}
\node[red,scale=1.5,anchor=center] at (2,0+\boxespacing+0.15) {$\cdot$};
\node[red,scale=0.5,anchor=east] at (2,0+\boxespacing+0.15) {$W(\TT)W^{*}(\TT)$};
\node[red,scale=1.5,anchor=center] at (2,0-\boxespacing-0.15) {$\cdot$};
\node[red,scale=0.5,anchor=west] at (2+0.05,0-\boxespacing-0.15) {$W^{T}(\TT)W^{\dagger}(\TT)$};

\node[anchor=west] at (2+\boxespacing+0.6,-0.01) {$=$};
\begin{scope}[xshift=0cm]

\Mbox{4}{0}{M}
\node[red,scale=1.4,anchor=center] at (4,0+\boxespacing+0.15) {$\cdot$};
\node[red,scale=0.5,anchor=east] at (4,0+\boxespacing+0.15) {$\tilde{F}$};
\node[red,scale=1.4,anchor=center] at (4,0-\boxespacing-0.15) {$\cdot$};
\node[red,scale=0.5,anchor=west] at (4,0-\boxespacing-0.15) {$\tilde{F}^{\dagger}$};
\end{scope}

\end{tikzpicture}
\end{equation}
where $\tilde{F}=W(F)\ee^{-\ii \frac{\pi}{2} (m^{(1)}+m^{(2)})^2}$. 
The relation between $V(\TT)$ and $\iota(\theta)$ can also be derived as
\begin{equation}
    \begin{tikzpicture}
\def\Mbox#1#2#3{
    \draw (#1-\boxespacing,#2-\boxespacing) -- ++(0,2*\boxespacing) -- ++(2*\boxespacing,0) -- ++(0,-2*\boxespacing)-- cycle;
    \node[anchor=center] at (#1,#2) {$#3$}; 
    \draw[thin] (#1,#2+\boxespacing)-- ++(0,0.35);
    \draw[thin] (#1,#2-\boxespacing)-- ++(0,-0.25);
    \draw[thin] (#1-\boxespacing,#2)-- ++(-0.25,0);
    \draw[thin] (#1+\boxespacing,#2)-- ++(0.25,0);
    \draw [my dash,thin] (#1+0.1,#2+\boxespacing)-- ++(0,0.35);
    \draw [my dash,thin] (#1+0.1,#2-\boxespacing)-- ++(0,-0.25);
    \draw[thin] (#1,#2+\boxespacing+0.35) arc (-180:-360:0.05);
    \draw[thin] (#1,#2-\boxespacing-0.25) arc (-180:0:0.05);
}

\Mbox{0}{0}{M}
\node[red,scale=1.5,anchor=center] at (0,0+\boxespacing+0.25) {$\cdot$};
\node[red, scale=0.6, anchor=west] at (0,0+\boxespacing+0.35) {$\ee^{\ii\theta m^{(1)}_{\lambda}}$};
\node[red,scale=1.5,anchor=center] at (0,0+\boxespacing+0.15) {$\cdot$};
\node[red, scale=0.5, anchor=east] at (0,0+\boxespacing+0.15) {$W^{\dagger}(\TT)$};
\node[red,scale=1.5,anchor=center] at (0,0-\boxespacing-0.15) {$\cdot$};
\node[red, scale=0.5, anchor=west] at (0,0-\boxespacing-0.15) {$W(\TT)$};

\node[anchor=west] at (0+\boxespacing+0.25,-0.01) {$=\ee^{-\ii\theta}$};

\Mbox{2.2}{0}{M}
\node[red,scale=1.5, anchor=center] at (2.2,+\boxespacing+0.25) {$\cdot$};
\node[red,scale=0.6, anchor=west] at (2.2,+\boxespacing+0.35) {$\ee^{-\ii\theta m^{(1)}_{\lambda}}$};
\node[anchor=west] at (2.2+\boxespacing+0.25,0) {$=\ee^{-\ii\theta}$};
\Mbox{4.4}{0}{M}
\node[red,scale=1.5,anchor=center] at (4.4-\boxespacing-0.15,0) {$\cdot$};
\node[red, scale=0.5,anchor=north] at (4.4-\boxespacing-0.25,0)
{$\iota(-\theta)$};
\node[anchor=east] at (-0.5,-1.5) {$=$};
\Mbox{0.2}{-1.5}{M^{*}}
\draw[thin](0.2-\boxespacing-0.15,-1.5)--++(-0.15,0);
\draw[thin](0.2+\boxespacing+0.15,-1.5)--++(0.15,0);
\node[red,scale=1.5, anchor=center] at (0.2,-1.5+\boxespacing+0.25) {$\cdot$};
\node[red,scale=0.6, anchor=west] at (0.2,-1.5+\boxespacing+0.35) {$\ee^{\ii\theta m^{(1)}_{\lambda}}$};
\node[red, anchor=center, scale=1.5] at (0.2-\boxespacing-0.15,-1.5) {$\cdot$};
\node[anchor=south,red, scale=0.5] at (0.2-\boxespacing-0.25,-1.5) {$V(\TT
)$};
\node[anchor=south,red, scale=0.5] at (0.2+\boxespacing+0.25,-1.5) {$V^{\dagger}(\TT)$};
\node[red,scale=1.5,anchor=center] at (0.2+\boxespacing+0.15,-1.5) {$\cdot$};

\node[anchor=west] at (\boxespacing+0.4,-1.5) {$=$}; 
\Mbox{2.2}{-1.5}{M^{*}}
\draw[thin] (2.2-\boxespacing-0.25,-1.5)--++(-0.5,0);
\draw[thin]  (2.2+\boxespacing+0.25,-1.5)--++(0.3,0);
\node[red,scale=1.5,anchor=center] at (2.2-\boxespacing-0.4,-1.5) {$\cdot$};
\node[red,scale=0.5,anchor=north] at (2.2-\boxespacing-0.4,-1.5) {$V(\TT)$};
\node[red,scale=1.5,anchor=center] at (2.2-\boxespacing-0.2,-1.5) {$\cdot$};
\node[red,scale=0.5,anchor=south] at (2.2-\boxespacing-0.2,-1.5) {$\iota(\theta)$};
\node[red,scale=1.5,anchor=center] at (2.2+\boxespacing+0.25,-1.5) {$\cdot$};
\node[red,scale=0.5,anchor=north] at (2.2+\boxespacing+0.3,-1.5) {$V^{\dagger}(\TT)$};
\node[anchor=west] at (2.5+\boxespacing+0.35,-1.5) {$=$};
\begin{scope}[xshift=3cm]
\Mbox{1.8}{-1.5}{M}
\draw[thin] (1.8-\boxespacing-0.25,-1.5)--(1.5-\boxespacing-0.25,-1.5);
\draw[thin] (1.8-\boxespacing-0.25,-1.5)--++(-0.7,0);
\node[red,scale=1.5,anchor=center] at (1.8-\boxespacing-0.8,-1.5) {$\cdot$};
\node[red,scale=0.5,anchor=north] at (1.8-\boxespacing-0.75,-1.5) {$V(\TT)$};
\node[red,scale=1.5,anchor=center] at (1.8-\boxespacing-0.5,-1.5) {$\cdot$};
\node[red,scale=0.5,anchor=south] at (1.8-\boxespacing-0.5,-1.5) {$\iota(\theta)$};
\node[red,scale=1.5,anchor=center] at (1.8-\boxespacing-0.2,-1.5) {$\cdot$};
\node[red,scale=0.5,anchor=north] at (1.8-\boxespacing-0.3,-1.5) {$V^{\dagger}(\TT)$};
\end{scope}
\end{tikzpicture}
\end{equation}
where we used Eq.~\eqref{eq:tensor_eqs_fermion}. Thus, we have $V(\TT)\circ \iota(\theta)=\ee^{-\ii\theta} \iota(-\theta)$.

Like the bosonic case, the disorder parameter can also be calculated using the transfer matrix $T(\theta)\equiv\sum_{\gamma\delta}\left( \sum_{\alpha\beta}[W(\theta)]_{\beta\alpha} M_{\alpha\beta;\gamma\delta} \right) \vket{\gamma}\vbra{\delta}\equiv\vcenter{\hbox{\includegraphics[height=3.5em]{fig/trans_matrix.pdf}}}$
\begin{align}
\lrangle{U_A(\theta)}=\Tr([T(\theta)]^{\abs{\partial A}}) \approx  \sum_{l=1}^{k}[\lambda_{l}(\theta)]^{\abs{\partial A}}\,.
\label{}
\end{align}


Now we focus on the symmetries of $T(\theta)$. Similar to the  bosonic case, we have
\begin{equation}
    \begin{tikzpicture}
\def\Mbox#1#2#3{
    \draw (#1-\boxespacing,#2-\boxespacing) -- ++(0,2*\boxespacing) -- ++(2*\boxespacing,0) -- ++(0,-2*\boxespacing)-- cycle;
    \node[anchor=center] at (#1,#2) {$#3$}; 
    \draw[thin] (#1,#2+\boxespacing)-- ++(0,0.35);
    \draw[thin] (#1,#2-\boxespacing)-- ++(0,-0.25);
    \draw[thin] (#1-\boxespacing,#2)-- ++(-0.25,0);
    \draw[thin] (#1+\boxespacing,#2)-- ++(0.25,0);
    \draw [my dash,thin] (#1+0.1,#2+\boxespacing)-- ++(0,0.35);
    \draw [my dash,thin] (#1+0.1,#2-\boxespacing)-- ++(0,-0.25);
    \draw[thin] (#1,#2+\boxespacing+0.35) arc (-180:-360:0.05);
    \draw[thin] (#1,#2-\boxespacing-0.25) arc (-180:0:0.05);
}

\node[scale=1,anchor=east] at (0-\boxespacing-0.15-0.3,0) {$V(\TT)T^{*}(-\theta)V^{\dagger}(\TT)=$}; 
\Mbox{0}{0}{M^*}
\node [red, scale=1.5, anchor=center] at (0,+\boxespacing+0.15) {$\cdot$};
\node [red,anchor=east,scale=0.6] at (0,+\boxespacing+0.15) {$W(\theta)$};
\node[anchor=west] at (0+\boxespacing+0.25,-0.01) {$=$};
\node[red,anchor=center,scale=1.5] at (-\boxespacing-0.15,0) {$\cdot$};
\node[red,anchor=north,scale=0.5] at (-\boxespacing-0.25,0) {$V(\mathcal{T})$};
\node[red,anchor=center,scale=1.5] at (+\boxespacing+0.15,0) {$\cdot$};
\node[red,anchor=north,scale=0.5] at (+\boxespacing+0.25,0) {$V^{\dagger}(\mathcal{T})$};
\Mbox{1.5}{0}{M}
\node [red, scale=1.5, anchor=center] at (1.5,+\boxespacing+0.25) {$\cdot$};
\node [red,anchor=east,scale=0.6] at (1.5,+\boxespacing+0.25) {$W(\theta)$};
\node [red, scale=1.5, anchor=center] at (1.5,+\boxespacing+0.1) {$\cdot$};
\node [red,anchor=west,scale=0.6] at (1.5,+\boxespacing+0.1) {$W^{\dagger}(\mathcal{T})$};
\node [red, scale=1.5, anchor=center] at (1.5,-\boxespacing-0.15) {$\cdot$};
\node [red,anchor=east,scale=0.6] at (1.5,-\boxespacing-0.15) {$W(\mathcal{T})$};
\node[anchor=west] at (1.5+\boxespacing+0.25,0) {$=$};
\Mbox{3}{0}{M}
\node [red, scale=1.5, anchor=center] at (3,+\boxespacing+0.25) {$\cdot$};
\node [red,anchor=east,scale=0.6] at (3,+\boxespacing+0.25) {$W(\theta)$};
\node [red, scale=1.5, anchor=center] at (3,+\boxespacing+0.1) {$\cdot$};
\node [red,anchor=west,scale=0.6] at (3,+\boxespacing+0.1) {$\Lambda^{(1)}(\theta)\Lambda^{(2)}(\theta)$};
\begin{scope}[xshift=0.8cm]
\node[anchor=west] at (3+\boxespacing+0.25,0) {$=$};
\Mbox{4.7}{0}{M}
\node [red, scale=1.5, anchor=center] at (4.7,+\boxespacing+0.25) {$\cdot$};
\node [red,anchor=east,scale=0.6] at (4.7,+\boxespacing+0.25) {$W(\theta)$};
\draw[thin] (4.7-\boxespacing-0.25,0)-- ++(-0.15,0);
\draw[thin] (4.7+\boxespacing+0.25,0)-- ++(0.15,0);
\node[red,scale=1.5,anchor=center] at (4.7-\boxespacing-0.2,0) {$\cdot$};
\node[red,scale=0.5,anchor=north] at (4.7-\boxespacing-0.2,0) {$\iota(\theta)$};
\node[red,scale=1.5,anchor=center] at (4.7+\boxespacing+0.3,0) {$\cdot$};
\node[red,scale=0.5,anchor=north] at (4.7+\boxespacing+0.3,0) {$\iota(-\theta)$};
\node[anchor=west] at (4.7+\boxespacing+0.4,0) {$=\iota(\theta)T(\theta)\iota(-\theta)$};
\end{scope}
\end{tikzpicture}
\label{eq:tensor_fermion_antiunitary}
\end{equation}
And
\begin{equation}
    \begin{tikzpicture}
\def\Mbox#1#2#3{
    \draw (#1-\boxespacing,#2-\boxespacing) -- ++(0,2*\boxespacing) -- ++(2*\boxespacing,0) -- ++(0,-2*\boxespacing)-- cycle;
    \node[anchor=center] at (#1,#2) {$#3$}; 
    \draw[thin] (#1,#2+\boxespacing)-- ++(0,0.35);
    \draw[thin] (#1,#2-\boxespacing)-- ++(0,-0.25);
    \draw[thin] (#1-\boxespacing,#2)-- ++(-0.25,0);
    \draw[thin] (#1+\boxespacing,#2)-- ++(0.25,0);
    \draw [my dash,thin] (#1+0.1,#2+\boxespacing)-- ++(0,0.35);
    \draw [my dash,thin] (#1+0.1,#2-\boxespacing)-- ++(0,-0.25);
    \draw[thin] (#1,#2+\boxespacing+0.35) arc (-180:-360:0.05);
    \draw[thin] (#1,#2-\boxespacing-0.25) arc (-180:0:0.05);
}

\begin{scope}[xshift=-0.1]
\Mbox{0}{0}{M}
\draw[thin] (0-\boxespacing-0.15,0)--++(-0.2,0);
\draw[thin] (0+\boxespacing+0.15,0)--++(0.2,0);
\node [red, scale=1.5, anchor=center] at (0,+\boxespacing+0.15) {$\cdot$};
\node [red,anchor=east,scale=0.6] at (0,+\boxespacing+0.15) {$W(\theta)$};
\node[red,anchor=center,scale=1.5] at (-\boxespacing-0.25,0) {$\cdot$};
\node[red,anchor=north,scale=0.5] at (-\boxespacing-0.5,0) {$V(\TT)V^{*}(\TT)$};
\node[red,anchor=center,scale=1.5] at (+\boxespacing+0.25,0) {$\cdot$};
\node[red,anchor=south,scale=0.5] at (+\boxespacing+0.5,0) {$V^{T}(\mathcal{T})V^{\dagger}(\TT)$};
\end{scope}
\node[anchor=west] at (1+\boxespacing+0.1,-0.01) {$=$};
\begin{scope}[xshift=0cm]

\Mbox{2.5}{0}{M}
\node [red, scale=1.5, anchor=center] at (2.5,+\boxespacing+0.25) {$\cdot$};
\node [red,anchor=east,scale=0.6] at (2.5,+\boxespacing+0.25) {$W(\theta)$};
\node [red, scale=1.5, anchor=center] at (2.5,+\boxespacing+0.1) {$\cdot$};
\node [red,anchor=west,scale=0.6] at (2.5,+\boxespacing+0.1) {$\tilde{F}$};
\node [red, scale=1.5, anchor=center] at (2.5,-\boxespacing-0.15) {$\cdot$};
\node [red,anchor=east,scale=0.6] at (2.5,-\boxespacing-0.15) {$\tilde{F}^{\dagger}$};

\node[anchor=west] at (2.5+\boxespacing+0.4,0) {$=T(\theta)$};
    
\end{scope}
\end{tikzpicture}
\end{equation}

Thus, the symmetries of $T(\theta)$ can be summarized as follows:
\begin{align}
    Q\iota(\theta)V^{*}(\TT) \circ T(\theta)&=T(\theta)\nonumber\\
    V(\TT)V^{*}(\TT)\circ T(\theta)&=T(\theta).
\end{align}
where we use Eq.~\eqref{eq:T_hermicity}. One can further check $Q V^{*}(\TT)=V(\TT)Q^{*}$ like the bosonic case. Thus,  $[Q\iota(\theta)V^{*}(\TT)]^2=\ee^{-\ii\theta}V(\TT)V^{*}(\TT)$. By absorbing the phase factor, we have 
\begin{align}
\VV_{\theta}\equiv \ee^{\ii \frac{\theta}{2}}Q\iota(\theta) V^{*}(\TT)\,, \quad \VV_{\theta+2\pi}=-\VV_\theta,\quad \VV_{\theta}^2=V(\TT) V^{*}(\TT)\,, \quad [\VV_{\theta},T(\theta)]=0 
\end{align}

Now let's investigate the spectral flow of $T(\theta)$. Let $v^r_0(0)$ be the unique (right) dominant eigenstate of $T(0)$, we have
\begin{align}
    V(\TT)\KK \cdot v^{r}_0\sim v^{r}_0\quad \Rightarrow \quad V(\TT)V^{*}(\TT)\cdot v^r_0=v^r_0\,,\quad \VV_{0}\cdot  v^{r}_0=s_0 v^r_0
\end{align}
where $s_0=\pm1$. As $\theta$ flows $0$ to $2\pi$, continuity implies that 
\begin{align}
    V(\TT)V^{*}(\TT)\cdot v^r_0 (2\pi)=v^r_0(2\pi)\,,\quad \VV_{0} \cdot v^{r}_0(2\pi)=-s_0 v^r_0(2\pi)
\end{align}
Like the bosonic case, the change of quantum number indicates the crossing of the magnitude of two eigenvalues, and hence the non-smooth behavior of $\abs{\lrangle{U_{A}(\theta)}}$.

Similar to the bosonic case, when $\theta=0/\pi$, Eq.~\eqref{eq:tensor_fermion_antiunitary} implies an anti-unitary symmetry:
\begin{align}
    \iota(-n\pi)V(\TT)\KK\circ T(n\pi)=T(n\pi)
\end{align}
Due to the Kramers' theorem,  $v^r_0(\pi)$ and $\iota(-\pi)\cdot V(\TT)\cdot \KK \cdot v^r_0(\pi)$ are distinct states, given $V(\TT) V^*(\TT) \cdot v^{r}_0(0)=v^{r}_0(0)$ and the continuity of $v^r_0(\theta)$.

Thus, the simplest singularity is similar to the bosonic case, which is also presented in the main text. 

\subsection{Details of numerical studies}
To compute $\lag U_{A}(\theta)\rag$ in the Kane-Mele model with $N$ unit cells, we first exactly diagonalize the Hamiltonian in real space, obtaining $N$ single fermion eigenstates $\{\phi_{i}(r)\}$. The half-filled many-body wavefunction can be represented by a Slater determinant
\begin{align}
    \Phi(r_1,r_2,\cdots,r_{2N})=\frac{1}{\sqrt{(2N)!}}\begin{vmatrix}
        \phi_1(r_1) & \dots & \phi_{2N}(r_{1})\\
        \vdots & \ddots &\vdots\\
        \phi_{1}(r_{2N}) & \cdots & \phi_{2N}(r_{2N})
    \end{vmatrix}
\end{align}
Here, we pick $2N$ single fermion eigenstates with the lowest energy. To evaluate the disorder parameter, we introduce a transformation of $\phi_{j}(r)$:
\begin{align}
    \phi'_j(r)= \begin{cases}
        \ee^{\ii\theta}\phi_j(r), & \text{if}\ r\in A \\
        \phi_j(r), & \text{otherwise}
    \end{cases}\,.
\end{align}
Thus $\lag U_{A}(\theta)\rag$ is evaluated as follows
\begin{align}
    \lag U_{A}(\theta)\rag=\lag \Phi| \Phi'\rag= \frac{1}{(2N)!} \begin{vmatrix}
        \lag \phi_1|\phi'_1\rag & \dots &\lag \phi_1|\phi'_{2N}\rag\\
        \vdots & \ddots & \vdots\\
        \lag \phi_{2N}|\phi'_1\rag &\dots & \lag\phi_{2N}|\phi'_{2N}\rag
    \end{vmatrix}
\end{align}

\section{Noise Resilience of the protocol}
\label{app:noise}
Our proposed protocol exhibits robustness against a broad class of experimental noises.
We model noise through a a quantum channel $\mathcal{E}$ on the RDM: $\tilde{\rho}_A = \mathcal{E}[\rho_A]$.
This quantum channel is formulated as a finite depth local unitary (FDLU) $V$ acting on the system and environment where $\ket{\tilde{\psi}}=V(\ket{\psi}\otimes \ket{0}_{\rm{E}})$, and $\tilde{\rho}_{A} = \Tr_{\rm{E}}[\ket{\tilde{\psi}}\bra{\tilde{\psi}}]$
For a general experimental setting, $V$ commutes with the symmetry action on both system and environment, and therefore $\tilde{\rho}_{A}$ preserves the weak symmetry condition inherited from $\ket{\psi}$:
\begin{align}
    \ket{\psi}=U(g)\ket{\psi}\Rightarrow\tilde{\rho}_{A} = U_A(g) \tilde{\rho}_{A}U_A^{\dagger}(g)~,\quad
    \forall g\in G
\end{align}
Note that $\ket{\tilde{\psi}}$ belongs to the same correlated TI phase as $\ket{\psi}$, therefore the disorder parameter $\bra{\tilde{\psi}} \exp \{\ii \theta\, \tilde{n}_A\}\ket{\tilde{\psi}}$ with $\tilde{n}_A = n_{A} + n_{\rm{E},A}$ the total $U(1)$ number for both the system and the environment within region $A$, would exhibit similar non-smooth behavior.
However, as the environmental degrees of freedom are not experimentally accessible, the non-smooth signature in general fails.

If $V$ is \emph{strongly} $U(1)$ symmetric -- forbidding the charge exchange between the system and the environment, $\ket{\tilde{\psi}}$ then belong to the same SPT phase with a $U(1)$ symmetry $U(\theta) = \exp(\ii \theta\,\sum_j n_j)$, and the disorder parameter $\bra{\tilde{\psi}} U_A(\theta)\ket{\tilde{\psi}}$ retains the same non-analytic features as the noiseless case.
Let $p_{exchange}$ denote the probability of single-particle exchange noise per site, we can discard measurement outcomes with fluctuating total particle numbers. 
The resulting error scales as $p_{exchange}^2$, allowing successful detection when $p_{exchange}\ll 1$.

\bibliography{DisorderOperator}